\documentclass[12pt]{article}

\usepackage{hyperref}

\usepackage{appendix}
\usepackage{amsmath, amssymb, mathrsfs}
\usepackage{tikz, tcolorbox, color}
\usetikzlibrary{decorations.markings, arrows.meta, calc}
\tcbuselibrary{skins}
\usepackage {diagbox}
\usepackage{ulem}
\usepackage{physics}
\usepackage{fancybox}
\usepackage{fancyhdr}
\usepackage{bm}

\usepackage{epsf}
\usepackage{amsmath,amssymb}
\usepackage{latexsym}
\usepackage{graphicx,color}
\usepackage{wrapfig}
\usepackage{latexsym}
\usepackage{graphics}

\usepackage{subfigure}

\usepackage{color}
\usepackage{delarray,color} 

\usepackage[T1]{fontenc}
\usepackage{lmodern}
\usepackage[utf8]{inputenc}

\usepackage{mathtools}

\newcommand {\beq}{\begin{align}}
\newcommand {\eeq}{\end{align}}

\newcommand{\be}{\begin{equation}}
\newcommand{\ba}{\begin{align}}
\newcommand{\ea}{\end{align}}
\newcommand{\ee}{\end{equation}}

\newcommand{\beqa}{\begin{align}}
\newcommand{\eeqa}{\end{align}}

\newcommand{\unit}{\hbox to 3.8pt{\hskip1.3pt \vrule height 7.4pt
    width .4pt \hskip.7pt \vrule height 7.85pt width .4pt \kern-2.4pt
    \hrulefill \kern-3pt \raise 3.7pt\hbox{\char'40}}}

\def\matt[#1,#2,#3,#4]{\left(%
\begin{array}{cc} #1 & #2 \\ #3 & #4 \end{array} \right)}

\usetikzlibrary{arrows.meta,positioning,calc,decorations.pathmorphing,patterns}

\usepackage{tabularx}

\catcode`\@=11
\@addtoreset{equation}{section}

\catcode`@=12
\relax

\begin{document}

\begin{titlepage}

\setcounter{page}{0}

\renewcommand{\thefootnote}{\fnsymbol{footnote}}

\begin{flushright}

YITP-26-140,
RUP-26-29

\end{flushright}

\vskip 1.35cm

\begin{center}
{\Large \bf 
Realization of de Sitter spacetime in AdS/CFT
}

\vskip 1.2cm 

{\normalsize
Yasutaka Koga${}^{a}$~\footnote{yasutaka.koga(at)rikkyo.ac.jp}, 
Keiju Murata${}^{b}$~\footnote{murata.keiju(at)nihon-u.ac.jp} and 
Seiji Terashima${}^{c}$~\footnote{terasima(at)yukawa.kyoto-u.ac.jp}
}

\vskip 0.8cm

${}^{a}${\it
Department of Physics, Rikkyo University, 
Tokyo 171-8501, Japan}

${}^{b}${\it
Department of Physics, College of Humanities and Sciences, Nihon University, Sakura-josui,
 Tokyo 156-8550, Japan}

${}^{c}${\it
Center for Gravitational Physics and Quantum Information, \\
Yukawa Institute for Theoretical Physics, Kyoto University, Kyoto 606-8502, Japan
}

\end{center}

\vspace{12mm}

\begin{abstract}

We construct horizonless, everywhere-regular solitons containing a de~Sitter
core inside an asymptotically anti--de~Sitter spacetime, using an ordinary
action of Einstein gravity, a Maxwell field, and a charged scalar.  The object
is a gravitating gauged $Q$-ball whose false-vacuum interior acts as a positive
effective cosmological constant, sourcing an inner de~Sitter region joined
through a thin wall to a charged, asymptotically AdS exterior.  Two general
requirements are emphasized: the scalar must lie in a near-false-vacuum
(slow-roll--like) regime, 
and the exterior is driven to the extremal 
Reissner--Nordstr\"om--AdS locus in the thin-wall limit, in which the wall approaches the common would-be horizon.
Since the solution is static and
spherically symmetric, it may correspond to a semiclassical charge-$Q$ state
of the dual CFT, realizing de~Sitter space approximately within a holographic
theory.  We also discuss free fluctuations
whose de~Sitter--core modes appear as parametrically low-lying excitations above the state,
bulk reconstruction, and the
near-extremal regime, where finite-${\mathcal N}$ effects are expected to cut off the
formal classical throat and replace its independent free-field modes by an
interacting near-throat sector.  

\end{abstract}

\end{titlepage}

\newpage

\tableofcontents
\vskip 1.2cm

\setcounter{footnote}{0}

\section{Introduction and summary}

de~Sitter (dS) spacetime is of central importance in our understanding of the
physical world. Cosmological observations indicate a small positive
cosmological constant today, and the inflationary paradigm posits an early
epoch of approximately dS expansion; in both cases, the relevant geometry is
locally de~Sitter. A satisfactory theory of quantum gravity must therefore make
sense of dS space.

Yet quantum gravity in de~Sitter space remains poorly understood. Even the
basic question of what the Hilbert space is---its dimension, the meaning of the
horizon entropy, the role of the observer, and the absence of a clean
asymptotic boundary on which to anchor observables---is the subject of ongoing
debate, and there is no established nonperturbative formulation comparable to
what we have in the anti--de~Sitter (AdS) case. 
There, the AdS/CFT
correspondence \cite{Maldacena:1997re} provides a precise, nonperturbative definition of quantum
gravity in terms of an ordinary conformal field theory living on the boundary.

This contrast suggests a strategy: if one could realize de~Sitter space,
approximately, inside an asymptotically AdS spacetime, then the
holographic definition of the AdS bulk would give a handle on the dS region as
well, and questions about dS quantum gravity could be rephrased in the language
of a dual CFT.
\footnote{
The KKLT construction~\cite{Kachru:2003aw} also uses an AdS starting
point: a supersymmetric AdS vacuum is uplifted to a metastable
de~Sitter vacuum.  This differs from the present construction, which
retains asymptotically AdS boundary conditions and realizes a
de~Sitter region within a solitonic state.
}
What
would be especially valuable for the holographic question is a configuration
that is not a black hole but a horizonless, everywhere-regular solution:
on such a background the bulk theory is under control, and the dS region it
contains can be studied as part of an ordinary state of the dual CFT.

Several constructions of dS regions embedded in gravity have been proposed (if
one includes the asymptotically flat case as well), but each raises a difficulty
from the AdS/CFT viewpoint. Some support the dS region with a phantom scalar \cite{Bronnikov:2005gm},
whose wrong-sign kinetic term is unphysical. Others rely on nonlinear
electrodynamics, replacing the Maxwell term by a function of $F^2$ \cite{Ayon-Beato:1998hmi,Ayon-Beato:2000mjt,Bronnikov:2000vy,Dymnikova:2004zc}; this makes
the embedding in AdS/CFT delicate and tends to bring in ghost problems and the core is even inaccessible as hidden by the horizon. 
Still others consider explicitly time-dependent de~Sitter
bubbles~\cite{Alberghi:1999kd, Freivogel:2005qh} in AdS/CFT.
Identifying the dual CFT state and understanding how it encodes the
de~Sitter region is already challenging for static configurations without a horizon,
and time dependence makes this task more involved.

It is therefore important to realize a time-independent dS region
inside AdS using an ordinary action---standard Einstein gravity, a Maxwell
field, and a charged scalar with a conventional potential.
Recently,
Ogawa and Ishihara constructed a dS-core configuration in asymptotically flat
space~\cite{Ogawa:2023ive,Ogawa:2024joy}, and subsequently similar objects were found in
simpler models~\cite{Brihaye:2025dlq}. (We note, however, that what is described
as the dS-core region in the latter work appears, in our analysis, to actually
contain a black-hole horizon rather than to be horizonless.)
\footnote{Such a horizonless object with dS core is also known as {\it a gravastar} in a gravity community and the model constructed by Israel's junction method~\cite{Mazur:2004fk} has been extensively investigated.
See, e.g., Ref.~\cite{Cardoso:2019rvt} for its astrophysical property and Refs.~\cite{Chen:2025cee,Chen:2025ywj,Chen:2025jbf} for works in the AdS/CFT context.}

\medskip
In this paper we show that there exist horizonless, regular dS-core solitons in
asymptotically AdS space, built from such an ordinary action, and we
discuss how they are to be interpreted in AdS/CFT. Concretely, the object is a
gravitating gauged $Q$-ball: a charged scalar condensate whose interior sits in
a false vacuum, so that the matter behaves as a positive effective cosmological
constant and sources an inner de~Sitter region, while the exterior is a charged,
asymptotically AdS geometry. The two are joined by a thin matter wall, and the
full configuration carries no horizon and no singularity.\footnote{
Here, considering an asymptotic AdS space can also be understood as putting gravity into a 'box' to eliminate IR divergences, thereby making the mathematics more tractable. Furthermore, the Hilbert space of the realized approximate dS space is non-trivial, even though an external observer is not included.
} 

We first set up the model and the field equations
(Sec.~\ref{sec:model}), and identify the conditions under which a large
de~Sitter--like core forms (Sec.~\ref{sec:dscore}). Two of
these conditions are worth emphasizing, because they are not special to AdS but
hold for a dS core in gravity quite generally. First, the core exists only when
the scalar sits in a near-false-vacuum (slow-roll--like) regime, where the
potential energy dominates the frequency and gradient terms; this is what makes
the interior effectively a cosmological constant. 
This requires the effective potential of the scalar to be almost flat in some region.
Second, in the thin-wall limit of such a de~Sitter core, horizonlessness and regularity force its exterior to be near-extremal: the smooth thin wall cannot change the metric function by an order-one amount, so the exterior must be cut very close to---and on the regular side of---an extremal horizon.
We give a coordinate-free argument for
this, and corroborate it with explicit numerical solutions
(Sec.~\ref{sec:dS-core-RN-exterior}), exhibiting
representative regular solutions with a de~Sitter core, a thin wall, and a
near-extremal Reissner--Nordstr\"om--AdS exterior.\footnote{
When the approximately de~Sitter core considered here is realized,
its radius satisfies \(H^{-1}\simeq r_{\rm e}\), where \(r_{\rm e}\)
is the corresponding extremal RN--AdS radius
(see Sec.~\ref{sec:fuzzball-buchdahl}).
Thus, the core cannot be made arbitrarily large at fixed asymptotic
mass and charge in a given theory. Its associated entropy scale is
of order \(S_{\rm BH}\), rather than an independently adjustable,
arbitrarily large interior entropy at fixed exterior parameters.
In this sense, the construction avoids the bag-of-gold problem \cite{Wheeler1964, tHooft:1993dmi, Marolf:2008tx}.
}

We then turn to the holographic interpretation (Sec.~\ref{sec:adscft}). The
solution has the symmetries of a static, spherically symmetric state, 
so it may correspond to a semiclassical
charge-$Q$ state 
of the dual CFT
on the sphere; in
this sense de~Sitter space is realized, approximately, within a holographic
theory. 
Because the core sits at large gravitational redshift, $\sigma_{\rm in}\ll1$,
its excitations appear in the CFT at parametrically small energies above the
state.  The de~Sitter--core physics is thus encoded in the low-energy
excitations of the CFT above this heavy charge-$Q$ state.
We also discuss the near-extremal regime, in which finite-\(\mathcal N\)
effects are expected to cut off the free semiclassical throat at the
crossover scale \(T_H r_h\sim S_{\rm BH}^{-1}\) and replace its
independent low-energy modes by interacting near-throat dynamics.
In the classical thin-wall limit, where the wall thickness becomes
negligible compared with the de~Sitter radius, the de~Sitter and
RN--AdS stretched surfaces coincide, suggesting a possible relation
to de~Sitter static-patch holography.

Finally, in Appendix~\ref{sec:OI} we formulate an asymptotically AdS version of
the two-field Ogawa--Ishihara model. 
We present the
action and the corresponding equations of motion; we do not, however, solve them
numerically here, and a study of its solutions is left for future work.

Throughout, our metric signature is $(-,+,+,+)$, and we work in units where $c=\hbar=1$.

\section{Models for de Sitter core in asymptotic AdS}
\label{sec:model}

\subsection{Action and scalar potentials}

In this paper, we consider four-dimensional Einstein--Maxwell--Higgs theory with a negative cosmological constant $\Lambda=-3/L^2$:
\begin{align}
S
=\int d^4x\,\sqrt{-g}\Bigg[
\frac{1}{16\pi G}\,(R-2\Lambda)
\;-\;\frac{1}{4}F_{\mu\nu}F^{\mu\nu}
\;-\;(D_\mu\phi)^{\ast}(D^\mu\phi)
\;-\;V(|\phi|)
\Bigg],
\label{eq:action}
\end{align}
where
\begin{align}
F_{\mu\nu}=\partial_\mu A_\nu-\partial_\nu A_\mu,
\qquad
D_\mu\phi=\partial_\mu\phi+i e\,A_\mu\,\phi.
\end{align}
We take $e>0$. 
(Here, the overall coefficient of the kinetic term of the scalar is different from the usual choice by a factor $2$.)

Throughout this paper, we mainly consider the following exponential scalar potential:
\begin{align}
V(|\phi|)
= \frac{m^2}{\eta^2} (1-e^{ -\eta^2 |\phi|^2}),
\label{eq:potential2}
\end{align}
which was considered in \cite{Brihaye:2025dlq}, as well as potentials that behave as
\(V(|\phi|)=m^2|\phi|^2+\cdots\) for sufficiently small \(|\phi|^2\). 
In this paper, we numerically construct globally regular, horizonless solutions for this potential.\footnote{
This potential does not have a false vacuum in the strict sense, but $|\phi|=\infty$ is formally regarded as a false vacuum. Below, the false vacuum for this potential means this point.
}

In the context of $Q$-balls, the following sextic self-interaction potential is widely used:
\begin{align}
V(|\phi|)
= m^2 |\phi|^2
- \lambda |\phi|^4
+ \kappa |\phi|^6,
\qquad
\lambda>0,\quad \kappa>0.
\label{eq:potential}
\end{align}
This potential is known to admit $Q$-ball solutions, namely localized configurations with nonvanishing $|\phi|$ carrying a conserved U(1) charge $Q$, even in flat spacetime. The choice $\lambda>0$ and $\kappa>0$ ensures that the potential is bounded from below while allowing an energetically favorable condensate at a finite field amplitude. In the present work, we do not construct numerical solutions for this sextic potential. Instead, we argue that our analysis provides evidence suggesting that globally regular solutions of the type considered here are unlikely to exist for this choice of potential.

\subsection{Ansatz for asymptotically AdS solutions with de Sitter cores}

Here we consider self-gravitating, electrically charged configurations whose matter sector is analogous to that of a gauged $Q$-ball. The complex scalar field $\phi$ rotates in the internal U(1) phase with frequency $\omega$, sourcing an electrostatic potential $A_0(r)$ while remaining spherically symmetric in space.\footnote{
As we will see later, the standard gauged $Q$-ball solution does not develop a de Sitter core. The solutions studied in this paper should therefore be regarded as $Q$-ball-like gravitating configurations rather than ordinary gauged $Q$-balls.
}
For a review of $Q$-balls, see Ref.~\cite{Heeck:2020bau}.

We work with the static, spherically symmetric metric:
\begin{equation}
  ds^{2}
  =-\sigma(r)^{2}N(r)\,dt^{2}
  +\frac{dr^{2}}{N(r)}
  +r^{2}d\Omega_{2}^{2},
  \qquad
  N(r)=1-\frac{2m(r)}{r}+\frac{r^{2}}{L^{2}}.
  \label{eq:metric}
\end{equation}
Here $m(r)$ is the usual mass function, $\sigma(r)$ is a lapse/redshift factor, and $L$ sets the AdS radius via $\Lambda=-3/L^2$. For the matter sector, we take the standard spherically symmetric gauged $Q$-ball ansatz:
\begin{equation}
  \phi(t,r)
  =
  f(r)\,e^{-i\omega t},
  \qquad
  A_{\mu}dx^\mu = A_0(r)dt\ .
  \label{eq:qball-ansatz}
\end{equation}
It is convenient to define the gauge-invariant effective frequency
\begin{align}
  \Omega(r)\equiv \omega - e\,A_0(r).
  \label{eq:OmegaDef}
\end{align}
Physically, $\Omega(r)$ plays the role of a chemical potential for the scalar condensate.

Regularity at the origin requires $f'(0)=0$ and $A_0'(0)=0$, i.e.\ the scalar profile is smooth and the electric field $E_r\sim -A_0'(r)$ vanishes at $r=0$.
Asymptotically, $f(r)\to 0$ so that we approach the trivial vacuum, and $A_0(r)$ falls off as a Coulomb potential in AdS.
The models with two gauge fields and two scalars considered in \cite{Ogawa:2023ive} and their gauged versions will only be discussed briefly in this paper, although the model itself is given in Appendix \ref{sec:OI}.

Because the scalar is complex and charged, the relevant conserved quantity is the electric/U(1) Noether charge
\begin{align}
Q_N=\int d^3x\,\sqrt{-g}\,j^0,
\qquad
j^\mu
=-i\,\Big[\phi^\ast (D^\mu\phi)-(D^\mu\phi)^\ast\phi\Big],
\label{eq:charge}
\end{align}
For the ansatz used below this charge becomes
\begin{equation}
 Q_N=4\pi\int_0^\infty dr\,\sigma r^2 j^t
 =8\pi\int_0^\infty dr\,\frac{r^2\Omega(r)f(r)^2}{\sigma(r)N(r)}.
 \label{eq:noether-charge-radial}
\end{equation}
The electric flux charge differs from $Q_N$ by the gauge coupling and sign conventions; the gauge-invariant quantity in the matter equations is $\Omega$.
This conserved charge supports the solitonic configuration.\footnote{
In contrast, in the SU(2) adjoint-Higgs monopole, stability is topological (magnetic charge $n\in\mathbb{Z}$).
Here it is non-topological: the object is sometimes called a (gauged) $Q$-ball.}

\subsection{Equations of motion}
\label{sec:eom}

Varying the total action \eqref{eq:action} with respect to the metric, gauge field and scalar field yields
\begin{align}
&G_{\mu\nu}+\Lambda g_{\mu\nu}
=8\pi G\,T_{\mu\nu}, \label{Estn}\\
&\nabla_\mu F^{\mu\nu}=e\,j^\nu \ ,\label{Aeq}\\
&D^2 \phi = \frac{1}{2}\frac{d V(|\phi|)}{d |\phi|}\frac{\phi}{|\phi|}\ ,\label{phieq}
\end{align}
where the energy momentum tensor is given by
\begin{equation}
T_{\mu\nu}=F_{\mu\rho}F_\nu{}^\rho - \frac{1}{4}g_{\mu\nu}F_{\rho\sigma}F^{\rho\sigma}
+2D_{(\mu} \phi^\ast D_{\nu)} \phi 
-g_{\mu\nu}(|D\phi|^2 + V(|\phi|))\ .
\end{equation}
We define the energy density, radial pressure, and tangential pressure by
$\varepsilon(r)\equiv -T^{t}{}_{t}$, $p_r(r)\equiv T^{r}{}_{r}$ and $p_t(r)\equiv T^{\theta}{}_{\theta}=T^{\phi}{}_{\phi}$. 
For the $Q$-ball ansatz \eqref{eq:qball-ansatz} with metric \eqref{eq:metric}, these quantities are explicitly given by 
\begin{align}
  &\varepsilon(r)
  = \frac{(A_0'(r))^{2}}{2\,\sigma(r)^{2}}
   + \frac{\Omega(r)^{2} f(r)^{2}}
          {\,\sigma(r)^{2}N(r)}
   + N(r)\,f'(r)^{2}
   + V\!\left(f(r)\right),
   \label{eq:energyQ}\\[2mm]
  &\varepsilon(r)+p_{r}(r)
  = 
  \frac{2 \Omega(r)^{2} f(r)^{2}}
          {\sigma(r)^{2}N(r)}
   + 2 N(r)\,f'(r)^{2},
   \label{eq:epsplusprQ}\\
   &\varepsilon(r)+p_{t}(r)
   =\frac{2 \Omega(r)^2 f(r)^2}{\sigma(r)^2 N(r)} + \frac{\Omega'(r)^2}{e^2\sigma(r)^2}
   \ .
   \label{eq:epsplusptQ}
\end{align}
Here and throughout, a prime denotes differentiation with respect to
the argument of the function; for example, $f'$ denotes differentiation
with respect to $r$, whereas $V'$ denotes differentiation with respect
to the scalar field.
The first two terms in $\varepsilon(r)$ are, respectively, the electric field energy density and the ``time-derivative''/rotation energy of the charged condensate.
The third is the spatial gradient energy, and the last is the potential energy. 
The Einstein equations~(\ref{Estn}) reduce to the first–order system as 
\begin{align}
  m'(r) &= 4\pi G\,r^{2}\,\varepsilon(r),
  \label{eq:mprime}\\[1mm]
  \frac{\sigma'(r)}{\sigma(r)} 
  &= \frac{4\pi G\,r}{N(r)}\Big(\varepsilon(r)+p_{r}(r)\Big) .
  \label{eq:sigmaprime}
\end{align}
From Eqs.~(\ref{phieq}) and (\ref{Aeq}), we obtain the matter field equations as
\begin{align}
  &\frac{1}{\sigma(r) r^{2}}
  \frac{d}{dr}\!\Big[\sigma(r) r^{2} N(r) f'(r)\Big]
  =
  -\frac{\Omega(r)^{2}}{\sigma(r)^{2} N(r)}\,f(r)
  + \frac{1}{2}\,V'\!\left(f(r)\right),
  \label{eq:feq-grav}\\
  &\frac{1}{\sigma(r) r^{2}}
 \frac{d}{dr}\!\left[\frac{r^{2}}{\sigma(r)} \Omega'(r)\right]
 = 2 e^2\,\frac{\Omega(r)\,f(r)^{2}}{\sigma(r)^{2} N(r)}.
 \label{eq:Omegaeq-grav}
\end{align}
The second equation corresponds to Gauss's law for the electric field sourced by the charged scalar condensate.\footnote{
In flat space ($\sigma=N=1$), \eqref{eq:Omegaeq-grav} reduces to the standard gauged $Q$-ball equation $\Omega''+\frac{2}{r}\Omega'=2 e^{2} f^{2}\Omega$.}

It is convenient to introduce the following dimensionless variables and parameters:
\begin{equation}
\begin{split}
&x \equiv m r,
\qquad
\mu(x) \equiv m\, m(r),
\qquad
h(x) \equiv \frac{f(r)}{m},
\\
&w(x) \equiv \frac{\Omega(r)}{m}, 
\qquad
\sigma(x) \equiv \sigma(r),\\
&\alpha \equiv 4\pi G m^2,
\qquad
\ell \equiv m L,\qquad  \tilde\eta \equiv \eta m\ ,\\
&\tilde{V}(h(x)) \equiv \frac{V(|\phi|)}{m^4} =\frac{1}{\tilde{\eta}^2}\Bigl(1-e^{-\tilde{\eta}^2 h^2}\Bigr),
\end{split}
\label{eq:dimless_defs1} 
\end{equation}
In terms of the dimensionless variables, the metric component $N(r)$ is written as
\begin{equation}
N(x) \equiv 1 - \frac{2\mu(x)}{x} + \frac{x^2}{\ell^2}.
\qquad
\label{eq:dimless_defs2} 
\end{equation}
The equations of motion in terms of the dimensionless variables are summarized as 
\begin{align}
&\mu'(x) = \alpha x^2\left[
\frac{w'(x)^2}{2e^2\sigma(x)^2}
+\frac{w(x)^2 h(x)^2}{\sigma(x)^2 N(x)}
+N(x)\,h'(x)^2
+\tilde{V}(h(x))
\right],
\label{eq:dS-mu}\\
&\frac{\sigma'(x)}{\sigma(x)} = \frac{\alpha x}{N(x)}\left[
\frac{2 w(x)^2 h(x)^2}{\sigma(x)^2 N(x)}
+2 N(x)\,h'(x)^2
\right],
\label{eq:dS-sigma}\\
&\frac{1}{\sigma(x)x^2}\frac{d}{dx}\!\left(\sigma(x)x^2N(x)h'(x)\right)
= \frac{1}{2}\tilde{V}'(h(x))
-\frac{w(x)^2}{\sigma(x)^2 N(x)}\,h(x),
\label{eq:dS-h}\\
&\frac{1}{\sigma(x)x^2}\frac{d}{dx}\!\left(\frac{x^2}{\sigma(x)} w'(x)\right)
=
\frac{2e^2\,w(x)\,h(x)^2}{\sigma(x)^2 N(x)}.
\label{eq:dS-w}
\end{align}
Here, we use the same convention as that stated below
Eq.~\eqref{eq:epsplusptQ}: a prime denotes differentiation with respect
to the argument of the corresponding function. For example, $\mu'$
denotes differentiation with respect to $x$, whereas
$\widetilde{V}'$ denotes differentiation with respect to $h$.
Eqs.~\eqref{eq:dS-mu}--\eqref{eq:dS-w}, together with
the definition of $N(x)$ in \eqref{eq:dimless_defs2}, are the correct
dimensionless first-order Einstein equations and second-order matter
equations for the spherically symmetric, self-gravitating, gauged $Q$-ball.

\subsection{Asymptotics and boundary data}

To obtain regular boundary conditions at the origin, we expand the fields around $x=0$ as
\begin{align}
\mu(x) &= \mu_1 x + \mu_2 x^2 + \cdots, \quad 
\sigma(x) = \sigma_0 + \sigma_1 x + \sigma_2 x^2 + \cdots, \nonumber\\
w(x) &= w_0 + w_1 x + w_2 x^2 + \cdots, \quad 
h(x) = h_0 + h_1 x + h_2 x^2 + \cdots .
\label{expand0o}
\end{align}
Here, the regularity of the spacetime requires $\mu(0)=0$. 
Substituting the above expansions into Eqs.~(\ref{eq:dS-mu})--(\ref{eq:dS-w}) and solving the equations order by order in powers of $x$, we obtain
\begin{align}
&\mu_1=\sigma_1=h_1=w_1=\mu_2=0, \quad
\sigma_2 = \frac{\alpha h_0^2 w_0^2}{\sigma_0}, \nonumber\\
&w_2 = \frac{e^2}{3} h_0^2 w_0, \quad
h_2 = \frac{h_0}{6\sigma_0^2} 
\left(
\sigma_0^2e^{-\tilde{\eta}^2 h_0^2}-w_0^2
\right).
\label{r0expand}
\end{align}
The coefficients $h_0$, $w_0$ and $\sigma_0$ remain free parameters of the near-origin expansion. All higher-order coefficients are likewise uniquely determined in terms of $h_0$, $w_0$ and $\sigma_0$. Since an overall constant factor of $\sigma$ can be absorbed into a rescaling of the time coordinate, we fix this freedom by imposing $\sigma_0=1$ during the numerical integration from the origin. After obtaining the complete solution, we rescale the time coordinate so that the metric takes the canonical form at spatial infinity, namely $\sigma(\infty)=1$.

Next, let us consider the asymptotic behavior of the solution at spatial infinity. The scalar field in an asymptotically AdS spacetime exhibits a power-law falloff with respect to the radial coordinate as
\begin{equation}
h(x)\simeq h_\infty\left(\frac{x_{\rm max}}{x}\right)^{\Delta_h}\ ,\quad 
\Delta_h=\frac{3+\sqrt{9+4\ell^2}}{2}\ ,
\label{eq:bc-h-Q}
\end{equation}
where $x_{\rm max}$ denotes the upper boundary of the numerical integration domain introduced for the numerical calculations later.
In the present work, we will fix the AdS radius to $\ell=100$, which gives $\Delta_h\simeq100$. Consequently, the regular scalar field decays extremely rapidly as $x$ increases. Because of this rapid decay, the backreaction of the scalar field on both the metric and the gauge field becomes negligible near spatial infinity. Therefore, the asymptotic solution is well approximated by the RN-AdS solution. The asymptotic behavior of the fields is thus given by
\begin{align}
&\mu(x) \simeq M-\frac{Q^2}{2x},  \quad 
w(x) \simeq w_\infty+\frac{c_w}{x}, \quad
\sigma(x) \simeq 1\ .
\label{expandinf2}
\end{align}
The coefficient $c_w$ is determined by substituting above expressions into the equations of motion with $h=0$, yielding
\begin{equation}
c_w=-\frac{eQ}{\sqrt{\alpha}}.
\end{equation}
We have used the freedom to rescale the time coordinate to impose the normalization condition $\sigma(\infty)=1$.
The constants $M$, $Q$, and $w_\infty$ are free parameters that can be specified at spatial infinity. Here, $M$ and $Q$ correspond to the dimensionless mass and electric charge of the spacetime, respectively. 
In terms of the original variables, $w_\infty$ is given by
\begin{align}
w_\infty=\frac{\omega-eA_0(\infty)}{m}.
\label{winf}
\end{align}
The chemical potential (or equivalently the external source for the conserved current in the dual CFT) is identified with the boundary value of the gauge potential, $A_0(\infty)$.\footnote{
$A_0(\infty)$ is the boundary value, which is the external physical parameter, as the chemical potential, in particular for the asymptotic AdS case.
Indeed, the "gauge" transformation that rotates the fields at the boundary should be regarded as a global transformation.
}
In the gauge where
$A_0(\infty)=0$, 
we simply have $w_\infty=\omega/m$, 
so that $w_\infty$ is the dimensionless oscillation frequency of the scalar field. In this gauge, the physical frequency $\omega$ can therefore be read off directly from the asymptotic value $w_\infty$.\footnote{
Instead, we can regard the given configuration with $ w_\infty$ as a theory with an arbitrary chemical potential $A_0(\infty)=\mu_A$ and $\omega =m w_\infty+ e \mu_A$.
This is because the Q-ball configurations are in a special form.
}

\section{de Sitter core in asymptotic AdS}
\label{sec:dscore}

A large-$Q$ gauged $Q$-ball typically exhibits the following structure:
\begin{itemize}
\item a nearly constant scalar amplitude $h(x)\simeq h_0$
in the interior,
\item a thin transition layer of thickness $\delta$, where $h$ rapidly falls to $0$,
\item an exterior region $h\simeq 0$ with a purely electric, Coulomb-like tail for $A_0$ (equivalently $w$).
\end{itemize}
In flat space this is the standard ``gauged $Q$-ball'' picture; with gravity and negative $\Lambda$ this lump backreacts on $N(r)$ and $\sigma(r)$, potentially producing a high-density core whose stress tensor can resemble an effective positive vacuum energy in the interior.
Further, if the matter is not excited in the interior region, the stress tensor is dominated by an effective positive vacuum energy, and the interior region would be dS spacetime.  

However, such a dS core cannot be realized in a generic Q-ball model, as we will see below.
To realize a dS core, the potential must satisfy a condition that we call the slow-roll condition.\footnote{
Even with this condition, which is satisfied by the potential \eqref{eq:potential2}, the existence of a dS-core solution is not guaranteed; we establish existence only numerically for part of the parameter space.
}
We will also show a simple geometric consequence of the thin-wall dS-core limit.  
If the dS-like region extends close to its static-patch horizon
and the metric function $N$ changes only slightly across the
thin wall separating the core from the RN--AdS exterior,
then the exterior must also begin at small $N$.
Regularity and horizonlessness then force that exterior to be near extremal.

In what follows, we focus on the local behavior of the matter fields
near the center, $r\simeq 0$, and derive the conditions required for
the formation of a dS-like core.
Consider integrating the equations of motion outward from $r=0$.\footnote{
Here \(\Omega(0)/\sigma(0)\) is the frequency measured in the local
proper time at the center.  One may temporarily choose a
center-normalized time coordinate with \(\sigma(0)=1\), but the
boundary-normalized core lapse used later is generally
\(\sigma_{\rm in}\ll1\).
}
For weak gauge and gravitational couplings,
$e\ll 1$ and $Gm^2\ll 1$, the effects of the gauge and gravitational
fields on the matter configuration are negligible over a sufficiently
small region around the center.
Likewise, the effect of the cosmological constant is locally negligible
in this region.
We can therefore study the leading local behavior by setting
$e=G=\Lambda=0$.
This approximation is intended only for the local analysis near the
center.
As the equations of motion are integrated toward larger $r$, the gauge
and gravitational backreaction can accumulate and eventually become
important, while the effect of the cosmological constant also becomes
relevant at sufficiently large distances.
Such global effects are outside the scope of the analysis below and
will be taken into account when constructing the full gravitating
solutions in Sec.~\ref{sec:dS-core-RN-exterior}.


\subsection{Minimal frequency for the global $Q$-ball without gravity}

Let us briefly review the standard argument for the existence of a minimum frequency in the case of a global $Q$-ball without gravity in Minkowski spacetime ($e=G=\Lambda=0$). In this limit, Eq.~(\ref{eq:feq-grav}) reduces to
\begin{equation}
f'' + \frac{2}{r} f'
=-\frac{dU_\omega(f)}{df}\ ,\quad 
U_\omega(f)\equiv 
\frac{1}{2}\left(\omega^2 f^2
-V(f)\right)\ .
\label{probeeq}
\end{equation}
This equation can be interpreted as the equation of motion for a particle moving in the potential $U_\omega$, with the radial coordinate $r$ playing the role of time and the term $(2/r)f'$ acting as a friction force.
Regularity at the origin and the requirement that the scalar field decay at spatial infinity impose the boundary conditions
\begin{equation}
f'(0)=0,
\qquad
f(r)\to0
\quad
(r\to\infty).
\end{equation}
In other words, the particle must start from rest at some point in the potential and asymptotically approach the origin after an infinitely long ``time'' $r$.
For this to be possible, there must exist some $f>0$ such that
$U_\omega(f)>0$. This immediately yields the following necessary condition on the frequency $\omega$:
\begin{equation}
\omega^2
>\omega_{\min}^2 \equiv  
\min_{f>0}
\frac{ V(f)}{f^2}\ .
\label{eq:omega-cond-shifted}
\end{equation}
We define $f_{\rm min}$ as the value of $f$ that minimizes
$V(f)/f^2$. By definition, it satisfies
\begin{equation}
    V(f_{\rm min}) = \omega_{\rm min}^2 f_{\rm min}^2,
    \qquad
    V'(f_{\rm min}) = 2\omega_{\rm min}^2 f_{\rm min},
    \label{VVp}
\end{equation}
where the second relation follows from the stationarity condition $\left. d(V(f)/f^2)/df \right|_{f=f_{\rm min}}=0$.
The corresponding results for the exponential potential~\eqref{eq:potential2} and the sextic potential~\eqref{eq:potential} are 
\begin{equation}
\omega_{\min}^2=
\begin{cases}
0 & (\text{Exponential potential})\\
m^2 - \frac{\lambda^2}{4\kappa} & (\text{Sextic potential})
\end{cases}
\ .
\label{eq:omega-min-shifted}
\end{equation}
Therefore, ungauged $Q$-balls exist only for
\begin{equation}
\omega_{\min}^2 < \omega^2 < m^2,
\label{omegacond}
\end{equation}
where $\omega^2<m^2$ guarantees exponential decay of the scalar field at large $r$, while
$\omega^2>\omega_{\min}^2$ prevents the configuration from spreading to
infinite radius. As $\omega\to\omega_{\min}$, the scalar field remains close to the value minimizing $V(f)/f^2$, i.e. $f_{\rm min}$, over an increasingly large interior region, while the transition layer near the surface stays finite. Consequently, the radius of the configuration diverges,
and the solution approaches the (Coleman) thin-wall limit.\footnote{
For this range, the Q-ball solution exists, even in the gauged Q-ball case with sufficiently small $e$ \cite{han2023existence}.}

\subsection{Requirements for a de Sitter core}
\label{s32}

We now list several necessary conditions for realizing a large dS-like core for the Q-ball in gravity,\footnote{
Such considerations are also done in \cite{Sakai:2007roh}.
} which may be regarded as a gravastar \cite{Mazur:2004fk}. 

In the probe limit, the energy density~(\ref{eq:energyQ}) reduces to
\begin{equation}
\varepsilon(r)
  = \omega^{2} f(r)^{2} 
   + f'(r)^{2} + V\!\left(f(r)\right),
\end{equation}
The second term represents the gradient contribution of the scalar field and can be neglected near the origin, since the regular expansion around $r=0$, Eq.~(\ref{r0expand}), gives
$f'(r)=\mathcal{O}(r)$.
Thus, near the center, for the above energy density to be approximately that of a positive cosmological constant, the potential energy must dominate over the contribution from the scalar rotation. Therefore, we require
\begin{align}
    V(f(0)) \gg \omega^2 f(0)^2\ .
\label{pc}
\end{align}

Furthermore, the scalar profile $f=|\phi|$ should remain approximately constant in the core region. Therefore, the scalar equation of motion, Eq.~(\ref{probeeq}), gives the balance condition
\begin{align}
    0 \simeq -\frac{dU_\omega}{df} = -\omega^2 f(0) +\frac12  V'(f(0))\ .
\label{f0}
\end{align}
Substituting the above relation into Eq.~(\ref{pc}) to eliminate $\omega$, we obtain 
\begin{align}
\delta\equiv \frac12 \left.  \frac{d \log V(f) }{d \log f} \right|_{f=f(0)} \ll 1.
\label{slc}
\end{align}
For example, if $V(f) \sim (f)^{2 a}$ near $f=f(0)$, this condition becomes $a \ll 1$.
Thus, this condition means that the potential is almost ``flat'' at $f=f(0)$, i.e., a kind of "slow-roll" condition.\footnote{
This is similar to the slow-roll condition for inflation, although it is not identical because the phase rotation $\omega$ is essential here.
Note also that this condition is analogous to, but distinct from, the usual
inflationary slow-roll condition.  Both favor a region in which the
potential is large while its logarithmic slope is small.  In
inflation this suppresses the slow-roll parameter and hence the
tensor amplitude, which is why plateau potentials such as
Starobinsky \cite{Starobinsky:1980te} or $\alpha$-attractor/T-model potentials \cite{Kallosh:2013hoa, Kallosh:2013yoa} are
observationally successful.  Here the same type of flattening has a
different role: it suppresses the rotational contribution of the
charged condensate relative to its potential energy, allowing the
core stress tensor to approach that of a positive cosmological
constant.
}
This will require fine-tuning of the potential. 

Next, to obtain the thin-wall structure, which has a large core region with approximately constant $f$, we require
$\omega^2-\omega_{\min}^2 \ll m^2$ for the global $Q$-ball in the absence of gravity. Motivated by this condition, we parameterize the central values of $\omega$ and $f(0)$ as
\begin{equation}
    \omega=\omega_{\min}+\delta\omega,
    \qquad
    f(0)=f_{\min}+\delta f,
\end{equation}
where $\delta\omega$ and $\delta f$ are assumed to be sufficiently small. $f_{\min}$ is defined below Eq.~(\ref{eq:omega-cond-shifted}).

However, we can show that the dS-like core in this model for the polynomial potential $V$ is almost impossible\footnote{
Here, we exclude the polynomial potential that contains very high-degree terms with finely tuned parameters, which would behave like a non-polynomial potential.
} 
because of \eqref{slc}.
Indeed, expanding the potential around $f=f_{\rm min}$ as
\begin{equation}
    V(f)
    = C_0 + C_1 (f-f_{\rm min})
    + C_2 (f-f_{\rm min})^2 + \cdots,
\end{equation}
the relations in \eqref{VVp} give
$C_0=\omega_{\rm min}^2 f_{\rm min}^2$ and
$C_1=2C_0/f_{\rm min}$.
Then, for $\omega_{\rm min}>0$, 
the slow-roll parameter $\delta$ defined in \eqref{slc} becomes
\begin{align}
    \delta \simeq \frac{1}{2} \frac{ C_1 +2 C_2\delta f}{C_0+C_1 \delta f} f_{\rm min} \simeq 1+ \left(\frac{2C_2}{C_1}- \frac{1}{f_{\rm min}}\right)\delta f\ ,
\end{align}
and for $\omega_{\min}=0$, we have $C_0=0$, and the above relation reduces to 
\begin{align}
    \delta \simeq 
    \frac{f_{\rm min}}{\delta f} , 
\end{align}
which cannot be much smaller than unity if $\delta f$ is a small perturbation.
Thus, even after including gravitational and gauge interactions, a solution with a dS core is not expected to exist for polynomial potentials.

On the other hand, for the non-polynomial potentials considered in this work, such as Eq.~\eqref{eq:potential2},
we can achieve the condition \eqref{slc}.
Indeed, for \eqref{eq:potential2}, 
$\omega_{\rm min}=0$ and $f_{\rm min}=\infty$, for which the expansion by $f-f_{\rm min}$ is not possible. For this case, we obtain,
\begin{align}
   \delta = \frac{\eta^2 f(0)^2}{e^{\eta^2 f(0)^2}-1}
  \simeq \frac{2\,\ln \frac{m}{\omega}}{\frac{m^2}{\omega^2}-1} \ .
\end{align}
Here, in the second equality, we have used the relation
$\omega^2 \simeq m^2 e^{-\eta^2 f(0)^2}$, which follows from Eq.~(\ref{f0}).\footnote{
For a finite core of size $R_{\rm core}$, the balance \eqref{f0} receives finite-size corrections of order $f(0)/R_{\rm core}^2$.  On the numerical solutions of Sec.~\ref{sec:dS-core-RN-exterior}, these corrections dominate the exponentially small potential slope, so $\Omega(0)$ is not literally given by this formula there.  The slow-roll condition \eqref{slc}, which depends only on the potential, is unaffected.
}
The above quantity is smaller than unity for $\omega \ll m$, making it possible to satisfy the slow-roll condition.\footnote{
The thin-wall case with $\omega_{\rm min}=0$ was considered in \cite{PaccettiCorreia:2001wtt}.
}

We can also directly examine the condition \eqref{pc}.
Using $f(0)=f_{\rm min}+\delta f$ and
$\omega=\omega_{\rm min}+\delta\omega$, we find
\begin{align}
    \frac{V(f(0))}{\omega^2 f(0)^2}
    \simeq
    \frac{V(f_{\rm min})}
         {\omega_{\rm min}^2 f_{\rm min}^2}
    \left(1-2\frac{\delta\omega}{\omega_{\rm min}}\right)
    =1-2\frac{\delta\omega}{\omega_{\rm min}} .
    \label{dsc}
\end{align}
Here, we have used
\begin{align}
    \frac{V(f(0))}{f(0)^2}
    =
    \frac{V(f_{\rm min})}{f_{\rm min}^2}
    +O(\delta f^2),
\end{align}
which follows from the definition of $f_{\rm min}$ as the minimum
of $V(f)/f^2$.
In the last equality of \eqref{dsc}, we have used \eqref{VVp}. 
Therefore, unless $\omega_{\rm min} \ll m$, we have $V(f(0))/(\omega^2 f(0)^2)\simeq 1$, which implies that the stress tensor is not dominated by an effective vacuum energy; correspondingly, $N \sigma^2 \simeq 1$ (i.e.\ $g_{tt}=-\sigma^2 N\simeq{\rm const.}$), and the core is not dS-like geometry.\footnote{
The relation \(N\sigma^2\simeq 1\) does not require \(N\) and
\(\sigma\) separately to remain close to unity.  Away from the wall,
the thin-wall interior is nearly homogeneous, \(f'(r)\simeq 0\), and
the scalar equation gives
\begin{equation*}
 \frac{\Omega(r)^2}{\sigma(r)^2N(r)}
 \simeq
 \frac{1}{2f(r)}\,V'(f(r)).
\end{equation*}
On the polynomial thin-wall branch, \(f(r)\simeq f_{\rm min}\), where
\(f_{\rm min}\) minimizes \(V(f)/f^2\).  Therefore,
\begin{equation*}
 \frac{1}{2f_{\rm min}}V'(f_{\rm min})
 =
 \frac{V(f_{\rm min})}{f_{\rm min}^2}
 =
 \omega_{\rm min}^2.
\end{equation*}
For weak gauging, \(\Omega(r)\simeq\Omega(0)\simeq\omega_{\rm min}\)
throughout the nearly homogeneous interior, and hence
\begin{equation*}
 N(r)\sigma(r)^2
 \simeq
 \frac{\Omega(r)^2}{\omega_{\rm min}^2}
 \simeq 1 .
\end{equation*}
Thus, \(N\) and \(\sigma\) may each vary substantially while their
product, and hence \(g_{tt}=-N\sigma^2\), remains approximately
constant.  This relation ceases to hold near the wall, where the
variation and gradient of \(f\) can no longer be neglected.
}
Thus, we need $\omega_{\rm min} \ll \delta \omega \ll m$ for the dS-like core. Note that in this case, the l.h.s. of \eqref{dsc} cannot be expanded in small $\delta \omega$.

Thus, the system can realize a dS-like core only if the scalar remains on the nearly flat part of the potential over a macroscopic region and then changes across a thin wall.  For the limiting geometric argument of Sec.~\ref{sec:fuzzball-buchdahl}, we impose two further properties of this thin-wall branch.  The dS-like region must extend close to its static-patch horizon on the inner side of the wall, and the wall must remain a smooth, controlled matter region whose integrated effect cannot change $\sqrt{N}$ by an order-one amount.  Under these conditions the RN--AdS side of the wall also lies at small $N$, and the exterior is forced to be near extremal.  These statements do not by themselves guarantee the existence of the branch; in Sec.~\ref{sec:dS-core-RN-exterior} we construct finite-distance representatives numerically.

\paragraph{Comments on Ogawa-Ishihara model.}
The model considered in \cite{Ogawa:2023ive} contains two complex scalar fields $\phi,\psi$ and a specific potential $V=\frac{\lambda}{4} (|\phi|^2-\eta^2)^2+\mu |\phi|^2 |\psi|^2$. 
For this model, if we integrate out $\phi$ by (classical) e.o.m., we will have an effective potential $V_{eff}(\psi)$ for $\psi$, which is of a non-polynomial type (and contains momentum-dependent parts).
For a large $|\psi|$, the effective potential $V_{\rm eff}(\psi)$ is expected to become a constant because a large $|\psi|$ gives an enormous mass to $\phi$ and then $\phi \simeq 0$ by the e.o.m., which implies $V_{\rm eff} \simeq \frac{\lambda}{4} \eta^4$.
Thus, this model will have properties similar to those of the model with \eqref{eq:potential2}.
Here, fine-tuning may be required for the potential of $\psi$, 
as $V(\psi) \simeq 0$; in particular, the mass term for $\psi$ should be zero or very small.

\paragraph{On the no-go theorem.}
In \cite{Sakai:2011wn}, it was shown that, for a $Q$-ball with
$e=0$ and $G=0$, a dS-like core cannot be realized.  Indeed, if we
define
\begin{equation}
E=(f')^2-V+\omega^2f^2,
\end{equation}
the equation of motion \eqref{eq:feq-grav} in this case gives
\begin{equation}
E'=-\frac{4}{r}(f')^2\leq0.
\end{equation}
Since $E|_{r=\infty}=0$, this implies $E(r=0)\geq0$, and hence, using
$f'(0)=0$,
\begin{equation}
V\bigl(f(0)\bigr)\leq\omega^2f(0)^2.
\end{equation}
Therefore, the potential energy cannot dominate at the center, as
would be required for a dS-like core.  Thus, this no-go theorem does
not exclude the ordinary flat-space $Q$-ball itself, but excludes the
realization of a dS-like core within it.

Coupling to gravity and the gauge field can change the monotonicity
property $E'\leq0$, so the above no-go argument does not directly apply
to the present system.  
In particular, if $\omega$ is small, the small gravitational correction can change the bound significantly.

It is important to note, however, that even
when a dS-like core is realized, the corresponding solution cannot be
continuously related to an ordinary $Q$-ball without gravity and the
gauge field.  The gravitational and gauge field effects are essential for the dS-like
core, and, as the gravitational and gauge couplings are weakened, the dS-like core
will 
disappear at some value of the coupling, 
or become infinitely large in the limit.\footnote{
These properties differ from those of the usual $Q$-ball models.  In this sense, one might say that the present solutions are not ordinary
$Q$-balls.
}
One might think that this theorem makes the analysis of subsection \ref{s32} redundant, since a $Q$-ball-like dS core continuously connected to $e=G=0$ is already excluded.
However, the slow-roll condition \eqref{slc}, which is invisible to this theorem, is also required, as we have seen.

\paragraph{Asymptotically AdS case.}
For the asymptotic AdS case, 
we further need the de Sitter-like condition.
We call a region ``dS-like'' if the local effective cosmological constant is
positive, i.e.
\begin{equation}
-\,\frac{3}{L^{2}} \;+\; 8\pi G\,\varepsilon_0 \;>\; 0
\quad\Longleftrightarrow\quad
8\pi G\,\varepsilon_0 \;>\; \frac{3}{L^{2}}.
\label{eq:dS-like-basic}
\end{equation}
Note that this is a necessary condition and not a sufficient condition for the core region to be approximately dS-like.
In addition, gradient, electric-field, and rotational contributions must be
subdominant and the scalar must remain approximately constant over a region
larger than \(H^{-1}\).
The core is
well approximated by a nearly constant scalar amplitude $f(0)$ and negligible
gradients/electric field,
and 
$\varepsilon_0 \simeq   \Omega(0)^2 f(0)^2+ V(f(0)) $.

\subsection{A thin wall forces a near-extremal exterior}
\label{sec:fuzzball-buchdahl}

We now isolate the simple geometric argument that will be used below.  Let
$r_{\rm w}^{-}$ and $r_{\rm w}^{+}$ denote, respectively, the inner edge of
the wall on the de~Sitter-core side and the outer edge on the RN--AdS side.
The argument requires two properties of the thin-wall dS-core limit.  First,
the dS-like region extends close to the static-patch horizon.  Second, the
wall is smooth and thin on the macroscopic scale, so that the metric function
cannot change by an order-one amount across it.  These two statements imply
that the RN--AdS exterior begins at parametrically small $N$.  Regularity and
horizonlessness then force the exterior to approach the extremal RN--AdS
locus.

Throughout this subsection we use the dimensionful areal radius $r$; in
particular, the scalar mass $m$ is not set to unity.  The relation to the
dimensionless variables of Sec.~\ref{sec:dS-core-RN-exterior} is
$x=mr$, $\ell=mL$, $M=m\mathcal{M}$, and $Q=m\mathcal{Q}$, where
$\mathcal{M}$ and $\mathcal{Q}$ are the dimensionful geometric mass and
charge parameters.
Note that the discussions in this subsection can be applied to the asymptotically flat case, i.e., $L=\infty$ also.

\paragraph{The two sides of the wall.}
In the core and in the exterior, respectively, the metric function is
approximated by
\begin{align}
  N_{\rm dS}(r)
  &= 1-H^{2}r^{2},
  &
  H^{2}
  &=\frac{8\pi G}{3}\,\varepsilon_{0}-\frac{1}{L^{2}}>0,
  \label{eq:dS-side-sec33}\\
  N_{\rm RN}(r)
  &=1-\frac{2\mathcal{M}}{r}
    +\frac{\mathcal{Q}^{2}}{r^{2}}
    +\frac{r^{2}}{L^{2}} .
  \label{eq:Next}
\end{align}
The first thin-wall condition is that the inner edge lies close to the dS
static-patch horizon,
\begin{equation}
  N_{-}
  \equiv
  N(r_{\rm w}^{-})
  \simeq
  1-H^{2}(r_{\rm w}^{-})^{2}
  \equiv
  \epsilon_{\rm dS}^{2},
  \qquad
  \epsilon_{\rm dS}\ll1 .
  \label{eq:dS-inner-edge-small}
\end{equation}
This is the geometric content of requiring the large, approximately constant
core to extend to a neighborhood of its would-be cosmological horizon.

The second condition is that the wall is a smooth thin region rather than an
independently prescribed distributional shell.  Introduce the proper radial
coordinate $d\rho=dr/\sqrt{N}$ and let $\rho_{\pm}$ correspond to
$r_{\rm w}^{\pm}$.  The Hamiltonian constraint is
\begin{equation}
  N'
  =
  \frac{1-N}{r}
  +\frac{3r}{L^{2}}
  -8\pi G r\,\varepsilon .
  \label{eq:constraint}
\end{equation}
Since $d\sqrt{N}/d\rho=N'/2$, it gives the exact relation
\begin{equation}
  \sqrt{N_{+}}-\sqrt{N_{-}}
  =
  \frac12
  \int_{\rho_-}^{\rho_+}d\rho
  \left[
    \frac{1-N}{r}
    +\frac{3r}{L^{2}}
    -8\pi G r\,\varepsilon
  \right],
  \qquad
  N_{+}\equiv N(r_{\rm w}^{+}).
  \label{eq:smooth-wall-junction}
\end{equation}
For a wall whose proper thickness is small compared with the macroscopic
radius, whose local stress tensor remains controlled by the fixed matter
scales, and which does not develop a distributional surface energy of order
$1/G$, the absolute value of the right-hand side is parametrically small.  We
denote its bound by
\begin{equation}
  \eta_{\rm w}
  \equiv
  \frac12
  \int_{\rho_-}^{\rho_+}d\rho
  \left|
    \frac{1-N}{r}
    +\frac{3r}{L^{2}}
    -8\pi G r\,\varepsilon
  \right|
  \ll1 .
  \label{eq:wall-small-change}
\end{equation}
In the ordinary smooth thin-wall scaling considered here,
$\Delta\rho_{\rm wall}/H^{-1}\to0$ while the local matter and curvature
scales remain finite, and hence $\eta_{\rm w}\to0$.

Equations~\eqref{eq:dS-inner-edge-small}--\eqref{eq:wall-small-change},
together with horizonlessness, imply
\begin{equation}
  0
  \leq
  \sqrt{N_{+}}
  \leq
  \epsilon_{\rm dS}+\eta_{\rm w},
  \qquad
  N_{+}
  =
  \mathcal{O}\!\left[
    (\epsilon_{\rm dS}+\eta_{\rm w})^{2}
  \right]
  \ll1 .
  \label{eq:outer-edge-small}
\end{equation}
Thus a smooth thin wall cannot convert an inner edge that is already close to
the dS horizon into an RN-side edge with $N=O(1)$.  The exterior necessarily
starts in a low-$N$ region.

\paragraph{Near-extremality of the exterior.}
Beyond $r_{\rm w}^{+}$ the scalar is negligible, so the geometry is
RN--AdS.  Consider first the superextremal side, which is the side realized by
the numerical solutions.  The RN--AdS function is everywhere positive and
has a global minimum at $r=r_*$,
\begin{equation}
  N'_{\rm RN}(r_*)=0,
  \qquad
  N_{\min}^{\rm RN}
  \equiv
  N_{\rm RN}(r_*)>0 .
  \label{eq:RN-positive-minimum}
\end{equation}
Since $r_*$ is the global minimum,
\begin{equation}
  0
  <
  N_{\min}^{\rm RN}
  \leq
  N_{\rm RN}(r_{\rm w}^{+})
  =
  N_{+}
  =
  \mathcal{O}\!\left[
    (\epsilon_{\rm dS}+\eta_{\rm w})^{2}
  \right].
  \label{eq:RN-min-bound}
\end{equation}
Therefore the minimum tends to zero while its first derivative already
vanishes.  The exterior approaches a double zero of $N_{\rm RN}$, namely the
extremal RN--AdS locus.  More explicitly, at fixed $\mathcal{Q}$ and $L$,
\begin{equation}
  \mathcal{M}_{\rm ext}(\mathcal{Q};L)-\mathcal{M}
  =
  \frac{r_{\rm e}}{2}\,N_{\min}^{\rm RN}
  +\mathcal{O}\!\left((N_{\min}^{\rm RN})^{2}\right),
  \label{eq:extremal-distance}
\end{equation}
where $r_{\rm e}$ is the extremal radius with the same $\mathcal{Q}$ and
$L$.  Hence the distance from the extremal locus is bounded by the small
quantity in Eq.~\eqref{eq:outer-edge-small}.  If
$L_2^{-2}\equiv\tfrac12N''_{\rm RN}(r_*)$, the same expansion gives
\begin{equation}
  |r_{\rm w}^{+}-r_*|
  =
  \mathcal{O}\!\left[
    L_2(\epsilon_{\rm dS}+\eta_{\rm w})
  \right].
  \label{eq:wall-near-throat}
\end{equation}
The minimum of the full numerical metric should be distinguished from
$N_{\min}^{\rm RN}$ unless the minimum lies in the scalar-free region.

The same endpoint is obtained if the formal RN--AdS parameters approach from
the subextremal side.  In that case the outer edge of the wall approaches the
formal outer horizon.  If the surface gravity remained nonzero, the exterior
extrinsic curvature would contain a term proportional to
$N'_{\rm RN}/\sqrt{N_{\rm RN}}$ and the static wall stresses would diverge.
A related result was obtained by Lemos and
Zaslavskii~\cite{Lemos:2007qbh} for quasiblack holes,
under different assumptions.
Thus the surface gravity must vanish, and the limiting zero is again double.

We conclude that
\begin{equation}
  \epsilon_{\rm dS}\to0,
  \qquad
  \eta_{\rm w}\to0
  \qquad
  \Longrightarrow
  \qquad
  \mathcal{M} \to
  \mathcal{M}_{\rm ext}(\mathcal{Q};L).
  \label{eq:extremal-limit}
\end{equation}
Charge is essential for this regular endpoint: a positive-mass
Schwarzschild--AdS exterior has no double zero.

\paragraph{The strict common-horizon limit and the numerical solutions.}
In the strict limit, the dS-side edge approaches $H^{-1}$ and the wall is
thin compared with the macroscopic radius.  Hence the two wall edges have
the same limiting areal radius in macroscopic units.  The RN--AdS side
approaches its extremal horizon at the same radius, so parametrically
\begin{equation}
  r_{\rm w}^{-}
  \simeq
  r_{\rm w}^{+}
  \simeq
  H^{-1}
  \simeq
  r_{\rm e} .
  \label{eq:common-horizon-limit}
\end{equation}
This is the common-horizon thin-wall limit used in the holographic discussion
of Sec.~\ref{sec:adscft}.  The explicit numerical solutions of
Sec.~\ref{sec:dS-core-RN-exterior} are finite-distance representatives: they
already have a near-extremal exterior, but the wall still lies inside the dS
radius and does not yet reach the strict limit~\eqref{eq:common-horizon-limit}.

\paragraph{The lapse function and the redshift factor $\mathcal{Z}$.}
The argument above concerns the spatial metric function $N$.  The remaining
Einstein equation~\eqref{eq:sigmaprime} together with Eq.~\eqref{eq:epsplusprQ} gives
\begin{equation}
  \frac{d\log\sigma}{d\rho}
  =
  \frac{4\pi G r}{\sqrt{N}}\,(\varepsilon+p_r)
  \geq0,
  \label{eq:logsigma}
\end{equation}
so $\sigma$ is monotonic outward and equals $1$ in the scalar-free exterior
with the boundary normalization.

On the numerical branch, 
$\sigma_{\rm in} \equiv \sigma(0) \ll 1$, and 
\begin{equation}
  \sigma_{\rm in}
  \gg 
  \sqrt{N_{\min}^{\rm RN}},
\end{equation}
which implies that 
${\mathcal Z}_{\rm min} \ll {\mathcal Z}_{\rm in} \ll 1$, where ${\mathcal Z}_{\rm in} ={\mathcal Z}(0)=\sigma_{\rm in}$.
The combination $\mathcal{Z}^2=\sigma^2 N$ reappears as
the potential prefactor governing the normal modes in
Sec.~\ref{sec:adscft}.

\paragraph{Relation to Buchdahl-type arguments.}
The ordinary Buchdahl theorem is not directly applicable: the configuration
is charged and anisotropic, and the potential-dominated core has
$p\simeq-\rho$.  The relevant general statement is instead the regularity of
a quasiblack-hole limit: a regular static surface can approach a would-be
horizon only if that horizon becomes extremal.  The argument above makes this
particularly transparent on the superextremal side, where the elementary
inequality~\eqref{eq:RN-min-bound} is already sufficient.  Thus the result is
consistent with Buchdahl-type compactness reasoning but does not rely on the
standard perfect-fluid Buchdahl bound.


\section{Numerical evidence for a horizonless de Sitter--core soliton with a Reissner--Nordstr\"om--AdS
exterior}
\label{sec:dS-core-RN-exterior}

In this section we exhibit an explicit, fully regular numerical solution of the
gravitating gauged $Q$-ball system with the exponential (non-polynomial)
potential \eqref{eq:potential2}, whose metric function $N(x)$ interpolates between three distinct
regimes: a de Sitter--like core, a soliton wall where the scalar
amplitude drops, and a Reissner--Nordstr\"om--anti--de~Sitter (RN--AdS)
exterior. The solution is horizonless, $N(x)>0$ everywhere, and the scalar
profile is a single smooth function that is localized without any artificial
matching at $h=0$.

\subsection{Numerical method}

We now solve Eqs.~(\ref{eq:dS-mu})--(\ref{eq:dS-w}) numerically.
Our strategy is to construct a global solution by matching the solution integrated outward from the origin ($x\simeq0$) with the solution integrated inward from spatial infinity ($x\simeq \infty$) at an intermediate point $x=x_{\rm mid}$. Both integrations are performed using the fourth-order Runge--Kutta method.
The boundary conditions near $r\simeq 0$ are given by
Eqs.~\eqref{expand0o} and \eqref{r0expand}, while those near
$r\simeq\infty$ are given by Eqs.~\eqref{eq:bc-h-Q} and
\eqref{expandinf2}.

The matching conditions at $x=x_{\rm mid}$ are given by
\begin{equation}
\begin{split}
&\mu(x_\text{mid}-0) = \mu(x_\text{mid}+0)\ ,\quad 
h(x_\text{mid}-0) = h(x_\text{mid}+0)\ ,\quad 
h'(x_\text{mid}-0) = h'(x_\text{mid}+0)\ ,\\
&\frac{\sigma(x_\text{mid}+0)}{\sigma(x_\text{mid}-0)}w(x_\text{mid}-0) = w(x_\text{mid}+0)\ ,\quad 
\frac{\sigma(x_\text{mid}+0)}{\sigma(x_\text{mid}-0)}w'(x_\text{mid}-0) = w'(x_\text{mid}+0)\ ,
\end{split}
\end{equation}
The discontinuity of $\sigma$ at $x=x_{\rm mid}$ can be absorbed by an appropriate rescaling of the time coordinate. Since the gauge potential $w$ is also rescaled under this transformation, the matching condition for $w$ includes the corresponding factor,
$\sigma(x_{\rm mid}+0)/\sigma(x_{\rm mid}-0)$
to ensure a consistent matching of the two solutions.
Since there are five matching conditions, they determine five of the six free parameters,
$h_0,\; w_0,\; M,\; Q,\; w_\infty,\; h_\infty$. 
Therefore, the solutions form a one-parameter family.
We use $h_0$ as the parameter labeling the family of solutions. Specifically, we fix a value of $h_0$ and solve the matching problem by the shooting method. We then repeat the procedure while gradually varying $h_0$, thereby generating the entire family of solutions.

The shooting procedure is implemented using the Newton--Raphson method. Since the asymptotic amplitude $h_\infty$ is extremely smaller than the other shooting parameters, directly treating $h_\infty$ as an independent variable leads to poor numerical stability. Instead, we use $\log_{10} h_\infty$ as the shooting parameter.

\subsection{Numerical solutions}

\begin{figure}
  \centering
\subfigure[]
 {\includegraphics[scale=0.6]{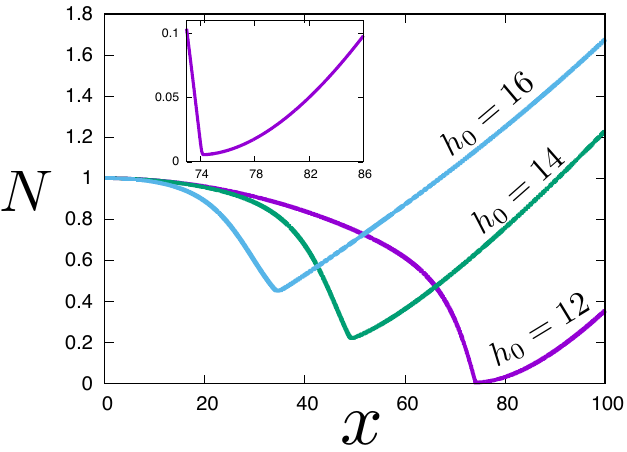}\label{N}
  }
  \subfigure[]
 {\includegraphics[scale=0.6]{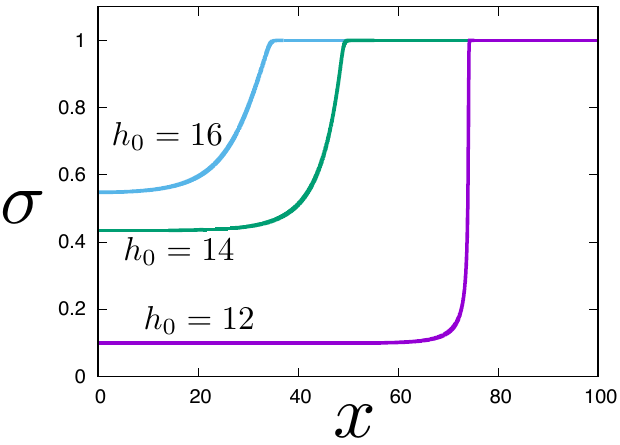}\label{sigma}
  }
   \subfigure[]
 {\includegraphics[scale=0.6]{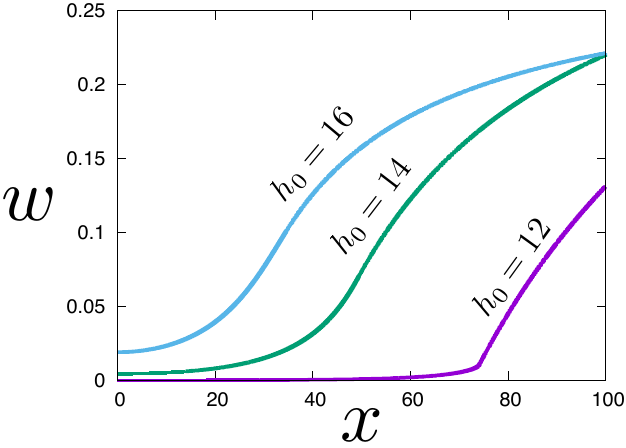}\label{w}
  }
  \subfigure[]
 {\includegraphics[scale=0.6]{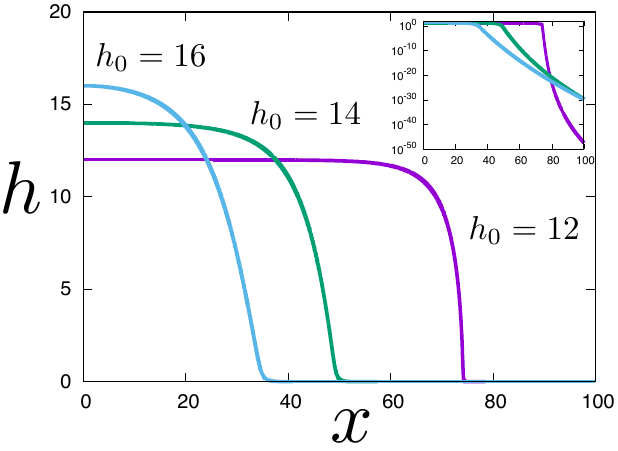}\label{h}
  }
 \caption{
 Numerical solutions for $N$, $\sigma$, $w$, and $h$. The parameters are fixed to the values given in Eq.~(\ref{eq:paramset}). The scalar field at the origin is varied as $h_0=12$, $14$, and $16$. Although the numerical integration is performed up to $x=200$, only the region $x\le100$ is shown for clarity.
 The inset in panel (a) shows a magnified view of the neighborhood of the local minimum for the $h_0=12$ solution. The inset in panel (d) shows the scalar field on a logarithmic scale.
}
 \label{sols}
\end{figure}

\begin{figure}
  \centering
  \subfigure[Energy density]
 {\includegraphics[scale=0.45]{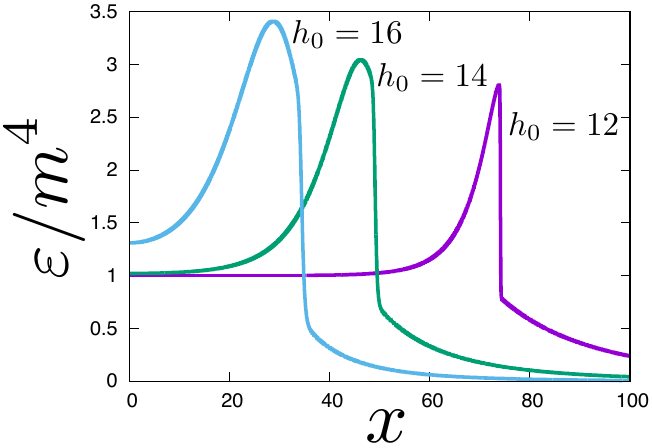}\label{energy}
  }
  \subfigure[Radial pressure]
 {\includegraphics[scale=0.45]{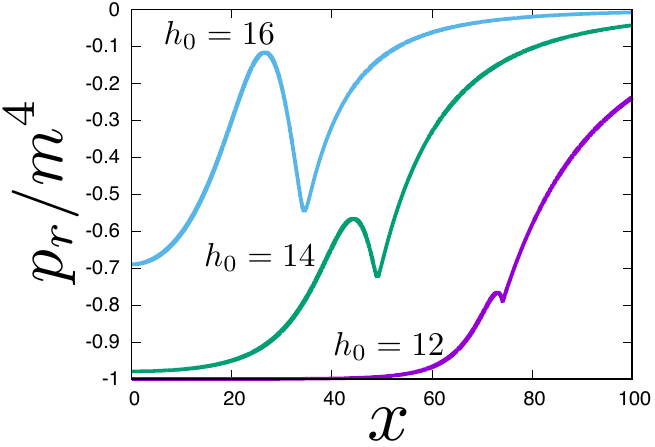}\label{pr}
  }
  \subfigure[Tangential pressure]
 {\includegraphics[scale=0.45]{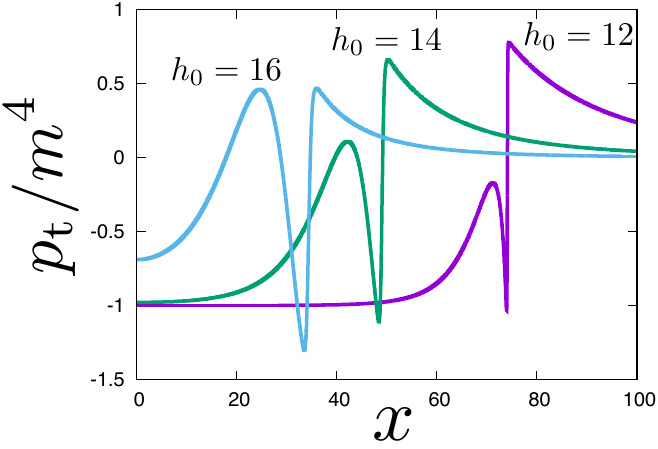}\label{pt}
  }
 \caption{Energy density, radial pressure, and tangential pressure. The parameters are fixed to the values given in Eq.~(\ref{eq:paramset}).  The scalar field at the origin is varied as $h_0=12$, $14$, and $16$.
}
 \label{energy_pressure}
\end{figure}

We fix the model parameters as 
\begin{equation}
  \tilde\eta=1,\quad
  \alpha=3\times10^{-4},\quad
  \ell=10^{2},\quad
  e=5\times10^{-3}\ .
  \label{eq:paramset}
\end{equation}
Figure~\ref{sols} shows the numerical solutions for $h_0=12$, $14$, and $16$. In all cases, the time coordinate has been rescaled so that $\sigma(\infty)=1$. To find this solution by the shooting method, the initial guesses for the shooting parameters
must be chosen sufficiently close to their true values. For readers who wish to reproduce our numerical calculation, we provide the converged values of these parameters for the case $h_0=12$ as
\begin{equation}
\begin{split}
&w_0=9.90019\times 10^{-4},\quad  M=153.976,\quad Q=119.763,\\ 
&w_\infty=0.476932,\quad \log_{10} h_\infty =-83.3784\ .
\end{split}
\label{paras}
\end{equation}
The value of $w_0$ is given in the normalization $\sigma(0)=1$ used for the outward integration (see below Eq.\eqref{r0expand}); in the boundary normalization it is smaller by the factor $\sigma(0)/\sigma(\infty)$.
The numerical domain is specified by
$x_{\rm min}=10^{-3}, x_{\rm mid}=75, x_{\rm max}=200$.
Starting from this solution, the entire solution family can be generated by gradually varying $h_0$. During this process, it is important to update the matching point $x_{\rm mid}$ accordingly. Specifically, at each step we choose $x_{\rm mid}$ to be the point at which   $h'(x)$ attains its minimum in the solution obtained at the previous step.
Figure~\ref{energy_pressure} shows the energy density, radial pressure, and tangential pressure corresponding to the solutions shown in Figure~\ref{sols}.
We define $x_{\rm wall}$ as the value of $x$ at which $h'(x)$ attains its minimum. 

These figures demonstrate that the obtained solutions indeed exhibit the following three characteristic regimes.
\paragraph{(i) de Sitter--like core.}
For $x\lesssim x_{\rm wall}$, where $x_{\rm wall}\simeq74,\;49,\;35$ for $h_0=12,\;14,\;16$, respectively, the metric component $N(x)$ behaves as
\begin{equation}
N(x)\simeq1-\frac{x^2}{\ell_{\rm dS}^2},
\end{equation}
indicating a de Sitter--like core. In this region, the scalar field sits in the false vacuum, $h\simeq h_0$, where the potential saturates, $V\to m^4/\tilde{\eta}^2$, yielding a nearly constant energy density, $\varepsilon_0\equiv\varepsilon(0)$ (see Fig.~\ref{pt}). From Eqs.~\eqref{eq:mprime} and \eqref{eq:dS-mu},
one finds that the de Sitter radius and the corresponding effective cosmological constant are given by
\begin{equation}
\ell_{\rm dS} = \left(\frac{2\alpha\varepsilon_{0}}{3m^4}-\frac{1}{\ell^{2}}\right)^{-1/2}\ ,\quad 
\Lambda_{\rm eff}=\frac{3m^2}{\ell_{\rm dS}^2}=8\pi G\varepsilon_{0}+\Lambda\ ,
\end{equation}
where the energy density at the origin is given by
\begin{equation}
\frac{\varepsilon_{0}}{m^4}=w_{0}^{2}h_{0}^{2}+\frac{1}{\tilde\eta^{2}} \bigl(1-e^{-\tilde\eta^{2}h_{0}^{2}}\bigr) .
\label{e0}
\end{equation}
The de Sitter radius predicted by the above expression agrees well with the numerical result.
Note that the values of $\varepsilon,\ p_r$, and $p_t$ at $x=0$ increases as $h_0$ increases toward $h_0=16$.
This implies the core deviates from the dS-like configuration $\varepsilon=-p_r=-p_t$ for larger $h_0$.

\paragraph{(ii) Soliton wall.}
Near $x=x_{\rm wall}$, the scalar field rapidly decreases from $h_0$ to a small value. Correspondingly, both the metric and the gauge field also undergo significant changes. In particular, the metric function $N(x)$ changes from the de Sitter--like behavior,
$N(x)\sim1-x^2/\ell_{\rm dS}^2$, 
to the asymptotically AdS behavior,
$N(x)\sim1+x^2/\ell^2$.
Although $N(x)$ becomes very small in the vicinity of the wall, it never vanishes. (For example, for the solution with $h_0=12$, the minimum value is
$\min N(x)\simeq5.5\times10^{-3}$.)
As shown in Fig.~\ref{energy}, the energy density is localized around $x\simeq x_{\rm wall}$. Although the solutions and their derivatives may appear to change discontinuously in this region, they are in fact smoothly connected.

\paragraph{(iii) RN--AdS exterior.}
For $x\gtrsim x_{\rm wall}$ the scalar is negligible and the geometry is governed
by the electric field and the cosmological term,
\begin{equation}
  N(x)\simeq 1-\frac{2M}{x}+\frac{Q^{2}}{x^{2}}+\frac{x^{2}}{\ell^{2}} .
  \label{eq:RNAdS}
\end{equation}


\begin{figure}
  \centering
  \subfigure[Mass–charge relation]
 {\includegraphics[scale=0.55]{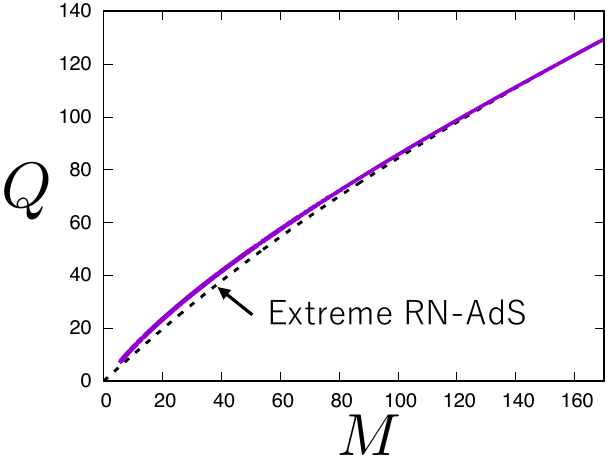}\label{MQ}
  }
  \subfigure[de Sitter radius and wall radius]
 {\includegraphics[scale=0.55]{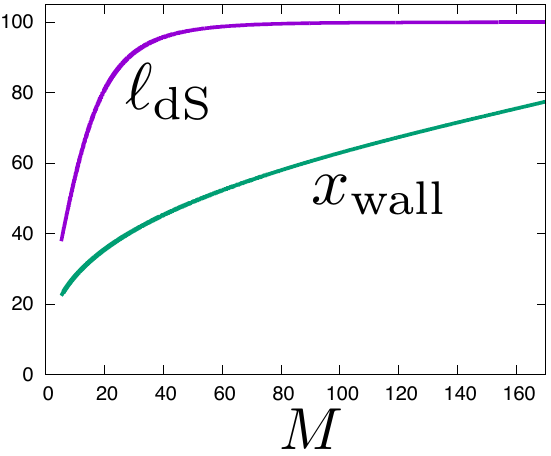}\label{Mlds}
  }
 \caption{The family of solutions obtained by varying $h_0$. Panel (a) shows the solutions in the $M$--$Q$ plane. Panel (b) shows the de Sitter radius $\ell_{\rm dS}$ and the wall position $x_{\rm wall}$ as functions of the mass $M$. The parameters are fixed to the values given in Eq.~(\ref{eq:paramset}). 
}
 \label{MQlds}
\end{figure}

Figure~\ref{MQlds} shows the family of solutions obtained with the model parameters fixed as in Eq.~(\ref{eq:paramset}). Panel (a) presents the $M$--$Q$ relation. The dashed curve corresponds to the extremal
Reissner--Nordström--AdS solution, whose mass is given by
\begin{equation}
M_{\rm ext}(Q)
=
\frac{\ell}{3\sqrt{6}}
\left(
2+\left(1+\frac{12Q^2}{\ell^2}\right)^{1/2}
\right)
\left(\left(1+\frac{12Q^2}{\ell^2}\right)^{1/2}-1
\right)^{1/2}\, .
\end{equation}
All solutions found in the present search lie in the superextremal region of the Reissner--Nordström--AdS parameter space. We emphasize, however, that this merely reflects the region in which our numerical search has been successful and does not necessarily imply the absence of regular solutions in the subextremal region. 

Panel (b) shows the de Sitter radius $\ell_{\rm dS}$ and the wall position $x_{\rm wall}$ as functions of the mass $M$. As the mass $M$ increases, the de Sitter radius $\ell_{\rm dS}$ approaches a constant value. This is because the contribution of the gauge--scalar coupling to the energy density near the center, represented by the $w_0^2h_0^2$ term in Eq.~(\ref{e0}), becomes negligible, while the energy at the top of the scalar potential dominates. Throughout the parameter range explored in this work, we find
$x_{\rm wall}<\ell_{\rm dS}$, indicating that the transition from the de Sitter core to the asymptotically AdS region occurs well inside the de Sitter radius.

Since the primary purpose of this paper is to demonstrate the existence of asymptotically AdS solutions with a de Sitter core and to discuss their implications for the AdS/CFT correspondence, we do not attempt an exhaustive search of the solution space. A more comprehensive exploration of the parameter space is left for future work.

We have focused on asymptotically AdS solutions. However, as shown in Ref.~\cite{Brihaye:2025dlq}, analogous solutions can also be constructed in the asymptotically flat limit, $\ell=\infty$. For completeness, we summarize the corresponding numerical solutions in Appendix~\ref{sec:dS-RN-largeL}.

\begin{figure}
  \centering
  \subfigure[Profiles of $h$ and $N$]
 {\includegraphics[scale=0.55]{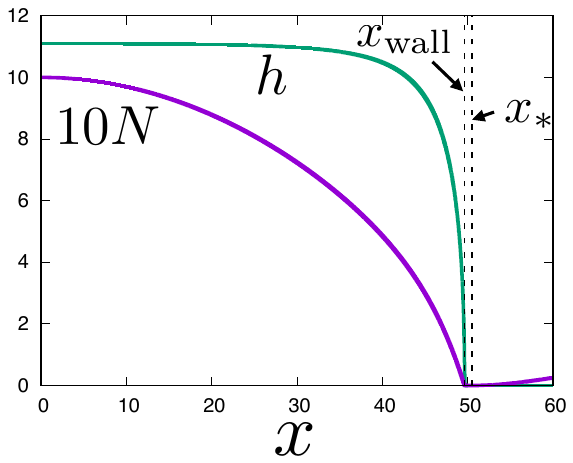}\label{handN}
  }
  \subfigure[Logarithmic zoom of $N$ near the throat]
 {\includegraphics[scale=0.55]{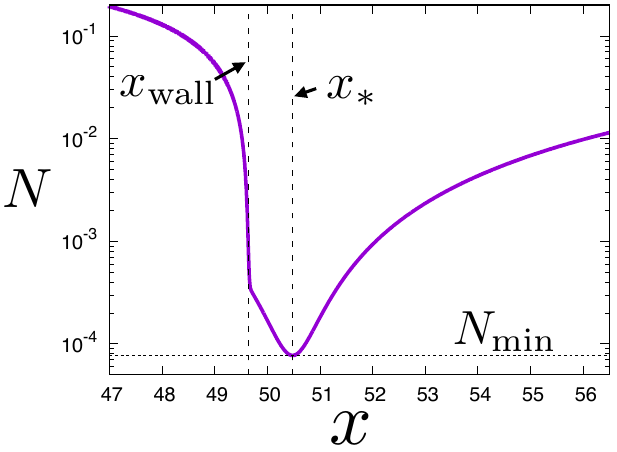}\label{Nzoom}
  }
 \caption{A near-extremal, small-gap horizonless soliton with the parameters
  $e=5\times10^{-3}$, $\alpha=4.52\times10^{-4}$, $\ell=\infty$, $h_0=11.1$,
  $w_0=5.50\times 10^{-3}$, $\tilde\eta=1$.
  (a) The scalar amplitude $h(x)$ (green) drops sharply across the thin wall while
  the metric function $N(x)$ (purple) develops a deep near-extremal minimum just
  outside the wall.
  (b) Logarithmic zoom of $N$ near the throat: the metric minimum
  $N_{\min}=7.7\times10^{-5}$ occurs at the would-be horizon $x_{*}=50.48$, only
  $|x_{\rm wall}-x_{*}|=0.84$ (i.e.\ $1.7\%$ of $L_2$) outside the wall at
  $x_{\rm wall}=49.64$.
}
 \label{fig:mingap}
\end{figure}

\paragraph{A near-extremal, small-gap representative.}
The wall approaches the would-be horizon as the
gravitational coupling $\alpha$ is increased and as the core amplitude $h_0$ is
lowered---one can construct explicit horizonless solitons whose wall sits very
close to the exterior would-be horizon while the exterior is driven deep into the
near-extremal regime.  Figure~\ref{fig:mingap} shows a representative example, for
the exponential potential and the parameter values
\begin{equation}
  \tilde\eta=1,\quad
  \alpha=4.52\times10^{-4},\quad
  \ell=\infty,\quad
  e=5\times10^{-3}\ ,
  \label{eq:paramset-mingap} 
\end{equation}
and 
\begin{equation}
\begin{split}
&h_0=11.1, \quad w_0=5.50436\times 10^{-3},\quad  M=50.4809,\quad Q=50.4828,\\ 
&w_\infty=0.241900,\quad \log_{10} h_\infty =-21.6840\ .
\end{split}
\end{equation}
For simplicity, we consider the asymptotically flat case, $\ell=\infty$, in this section. See Appendix~\ref{sec:dS-RN-largeL} for details of the asymptotic behavior at spatial infinity.
The resulting configuration is horizonless and superextremal ($M<Q$), with
\begin{equation}
 1-\frac{M}{Q}=3.80\times 10^{-5},\qquad
  N_{\min}=7.70\times10^{-5},\qquad \sqrt{N_{\min}}=8.77\times10^{-3}.
\end{equation}
The wall, located at $x_{\rm wall}\equiv m r_{\rm wall}=49.64$, 
sits just inside the metric minimum---the would-be
horizon---at $x_{*}\equiv m r_{*}=50.48$, so that
\begin{equation}
  |r_{\rm wall}-r_{*}|\,m=0.840,\qquad
  \frac{|r_{\rm wall}-r_{*}|}{r_{\rm wall}}=1.69\times10^{-2},\qquad
  \frac{|r_{\rm wall}-r_{*}|}{L_{2}}=1.66\times10^{-2},
\end{equation}
where $L_{2}$, defined by $L_{2}^{-2}=\tfrac12 N''(r_{*})$, is the near-horizon
AdS$_2$ radius, here $mL_{2}=50.70\simeq mr_{*}$.  The exterior is thus within
$N_{\min}\sim 7.7\times10^{-5}$ of extremality while the wall lies only a few percent
of the near-horizon scale inside the would-be horizon.  As anticipated, the de
Sitter core does not shrink in this limit: its radius remains of order
$\ell_{\rm dS}\sim1/H$ (here $x_{\rm wall}\sim50$), and it is the exterior throat,
not the core, that becomes deep.  Pushing $\alpha$ higher or $h_0$ lower continues
to reduce the gap, until the horizonless branch terminates at a Buchdahl-type
existence limit where the wall can no longer be supported outside the would-be
horizon; we have not resolved whether the gap reaches zero there or the branch
ends at finite gap.

\section{The de Sitter core in AdS/CFT}
\label{sec:adscft}

In this section we describe, within the AdS/CFT correspondence, the properties
of the (bulk) $Q$-ball-like de Sitter--core soliton with asymptotically AdS
boundary conditions. The construction of the solution itself, together with the
numerical evidence, has been given in the previous sections;
here we take such a solution as given and ask what it means
holographically.

\subsection{Why the soliton must be a gauged $Q$-ball}
\label{sec:why-gauged}

In AdS/CFT---and, more generally, in any consistent theory of quantum
gravity---it is widely believed that there are no global symmetries
\cite{Banks:2010zn}. 
The conserved charge $Q$ that stabilizes
the soliton therefore cannot be the Noether charge of a global $U(1)$; it must
be the charge of a gauged $U(1)$. This is why the relevant object is a
gauged $Q$-ball, with the scalar $\phi$ minimally coupled to a dynamical
Maxwell field $A_\mu$ as in \eqref{eq:action}, rather than the ungauged
(global) $Q$-ball familiar from flat-space field theory. The gauge field is not
an optional dressing: it is forced on us by the consistency of the bulk as a
quantum-gravitational system with a holographic dual, and it is precisely the
electric charge sourced by $A_0(r)$ that, as we have seen, allows the exterior
to approach extremality and thereby avoid a horizon.

\subsection{Structure of the solution and its regime of validity}
\label{sec:structure-validity}

The de Sitter--core soliton with asymptotically AdS boundary conditions is a regular solution. It has no curvature singularity and no horizon. Its
geometry consists of three regions:
\begin{itemize}
\item an inner region that is approximately a piece of de Sitter space
  (the scalar sits near a false vacuum $f\simeq f_0$, so that the matter acts as
  a positive effective cosmological constant and $N(r)\simeq 1-H^2 r^2$ with
  $H^2=(8\pi G/3)\,\rho_{\rm vac}-1/L^2$ set by the false-vacuum energy);
\item an outer region that is an asymptotically AdS, nearly extremal
  Reissner--Nordstr\"om black hole geometry---but without a horizon,
  since $N(x)>0$ everywhere and only comes close to a (double) zero;
\item a very narrow matter soliton (the wall) that joins the two,
  across which the scalar amplitude drops from $f_0$ to (essentially) zero.
\end{itemize}
The inner de Sitter region, the thin wall, and the outer Reissner--Nordstr\"om--AdS
region are glued together smoothly into a single horizonless, nonsingular
configuration.

As an effective field theory (EFT) solution, this geometry is trustworthy
provided the relevant length scales---most stringently, the thickness of the
matter soliton (the wall)---are sufficiently larger than the Planck length
$\ell_{\rm Pl}$. In that regime the curvatures and field gradients are
sub-Planckian everywhere and the classical (or semiclassical) description is
reliable. As the scales are reduced toward $\ell_{\rm Pl}$ the solution should
still retain qualitative meaning, but quantitative control is lost; we
return to this point in Sec.~\ref{sec:extremal-limit}, where the Planckian
regime turns out to be exactly the physically interesting one.

\subsection{Isometries, the dual Hamiltonian, and a dual state}
\label{sec:isometries-state}

The solution possesses two (commuting) Killing symmetries: time translation
$\partial_t$ and the $SO(d)$ rotations of the spatial sphere (here $d=3$, with
$SO(3)$ acting on the $S^{2}$; we keep $d$ general in the notation). Because
$\partial_t$ is a globally defined timelike Killing vector for this
asymptotically AdS geometry, one can define a conserved Hamiltonian $H$
generating translations in this time. Under the standard AdS/CFT dictionary,
this bulk Hamiltonian corresponds to the ordinary Hamiltonian of the dual CFT
quantized on the cylinder $\mathbb{R}_t\times S^{d-1}$. Consequently, the soliton
corresponds to a particular state of the dual CFT on a constant-time
slice $S^{d-1}$---an ordinary state in the ordinary Hilbert space, carrying the
electric charge $Q$ and the energy fixed by the bulk solution.\footnote{
The energy and electric charge are macroscopic ones, and the CFT state does not have definite energy and charge. 
}

It follows that de Sitter spacetime has been realized, approximately,
within quantum gravity---that is, within a holographic CFT. We emphasize the
sense in which this statement holds. We have assumed that a bulk action
admitting this solution can itself be realized as the bulk dual of some
holographic CFT. Since the action \eqref{eq:action} is of an entirely standard
type---Einstein gravity, a Maxwell field, and a charged scalar, with only a
somewhat special (but perfectly ordinary, sub-Planckian) scalar potential---this
assumption is a natural one. Granting it, the de Sitter region that actually
appears is not a global de Sitter universe: it is the static
patch, retained out to very close to (but not including) its cosmological
horizon, with the would-be horizon replaced by the thin wall and the exterior
AdS geometry. The dual CFT state is then a concrete, charge-$Q$ state that
encodes this static-patch de Sitter region.

We stress one open point. While the existence of a dual CFT state is guaranteed
by the dictionary, we have not identified which specific charge-$Q$ state
it is, nor have we translated the condition that produces the de Sitter core
(the false-vacuum/slow-roll condition on the scalar) into intrinsically CFT
language. We regard the CFT meaning of the de Sitter--core (false-vacuum)
condition as an important open problem.

\paragraph{Stability.}
The linear stability of the solutions remains to be established.  Since the
background is static, nonrotating, and \(SO(d)\)-invariant, linear
perturbations decompose into independent angular-momentum sectors.  Positive
angular-gradient terms suggest that the nonzero-\(\ell\) sectors are less
susceptible to tachyonic instabilities, while the spherically symmetric
\(\ell=0\) sector requires a direct analysis of the coupled scalar, Maxwell,
and gravitational perturbations.  We leave this stability analysis, as well
as the question of whether the solution is the lowest-energy state at fixed
charge, for future work.

\subsection{Free approximation around the soliton, and bulk reconstruction}
\label{sec:free-approx}

Let us first analyze free (quadratic) fluctuations around this background. We
begin with the case in which the length scale of the solution is small but
unrelated to the Planck length---parametrically $O({\mathcal N}^0)$ in the
't~Hooft-like counting, where ${\mathcal N}^2$ measures the number of degrees of freedom of
the CFT and $G/L^2 \sim 1/{\mathcal N}^{2}$. 
In this regime, the normal-mode
expansion and the free QFT approximation can be carried out in the ordinary way
(in practice, numerically). From it, bulk reconstruction---the rewriting of bulk
operators in terms of CFT operators---also proceeds in the ordinary way \cite{Banks:1998dd, Bena:1999jv, Duetsch:2002hc, Hamilton:2006az}.

A characteristic feature emerges when one uses the simple bulk
reconstruction picture \cite{Terashima:2021klf}, in which a bulk local operator is associated with
boundary data along the (null) directions reaching the AdS boundary. \footnote{
This approach is based on the outcomes discussed in \cite{Terashima:2017gmc, Terashima:2019wed, Terashima:2020uqu} within the operator formalism of AdS/CFT.
}
In this picture, an operator located
inside the de Sitter core is expressed in terms of CFT operators that are
very far away in time. The reason is geometric: the de Sitter core sits
behind the long, nearly extremal Reissner--Nordstr\"om throat, i.e.\ near the
(would-be) black hole horizon, so a signal from the core must traverse the
throat to reach the boundary. 
The light propagation time $\Delta t\sim \int dr/(\sigma N)$ accumulated across the throat grows large as
the geometry approaches extremality and the near-horizon region, pushing the boundary image of a core
operator to very late (and very early) CFT times.


It is worth assembling the free-field analysis into a single coherent picture. 

\paragraph{As normal modes, the core and the throat can be confined by the redshift.}
To display the geometric redshift effect explicitly, consider first a neutral,
minimally coupled probe scalar of mass \(\mu\).  Writing
\(\Phi=e^{-i\varpi t}Y_{jm}(\theta,\phi)u(r)/r\), its radial equation takes the
Schr\"odinger form
\begin{equation}
 \frac{d^2u}{dr_*^2}+\bigl[\,\varpi^2-V_j(r)\,\bigr]u=0,
 \qquad
 \frac{dr_*}{dr}=\frac{1}{\sigma N},
\end{equation}
with
\begin{equation}
 V_j(r)=\mathcal{Z}^2\left[
 \mu^2+\frac{j(j+1)}{r^2}
 +\frac{1}{r\sigma}\frac{d(\sigma N)}{dr}
 \right],
 \qquad
 \mathcal{Z}^2=\sigma^2N.
\label{eq:schrodinger-form}
\end{equation}
Throughout this subsection \(\varpi\) denotes the frequency conjugate to the
boundary time, i.e.\ the time normalized by \(\sigma(\infty)=1\).  The actual
scalar--Maxwell--gravity perturbations obey a coupled matrix equation; the
probe scalar is sufficient here to exhibit the redshift structure that is
relevant for localization.

Two features of \(\mathcal{Z}^2\) control the spectrum.  First, the wall and
throat are not themselves a positive potential barrier: there \(N\) is
parametrically small, so the effective potential is suppressed and the throat
forms a deep valley.  Second---and this is the essential point---the redshift
factor is not the same on the two sides of the wall, since \(\sigma\) is
sourced by the matter and rises across it.  Writing
\begin{equation}
 \sigma_{\rm in}\equiv\sigma(0)\ll1
\label{eq:redshift-step}
\end{equation}
for the total redshift between the center and the boundary, one has
\begin{equation*}
 \mathcal{Z}^2(r)\simeq
 \sigma_{\rm in}^2\bigl(1-H^2r^2\bigr)
\end{equation*}
in the de~Sitter-like core, with central value
\(\mathcal{Z}^2(0)=\sigma_{\rm in}^2\).  It decreases further toward the wall,
reaches \(\mathcal{Z}_{\rm min}^2\ll\sigma_{\rm in}^2\) in the near-wall/throat
region, and becomes \(\mathcal{O}(1)\) in the asymptotically AdS exterior.  At
fixed \(j\) and \(\mu\), the effective potential therefore exhibits a nested
redshift hierarchy.  This hierarchy is physical: a constant rescaling of the
time coordinate multiplies \(V_j\) and \(\varpi^2\) by the same factor
everywhere and cannot remove their relative steps.

The consequence is a nested pair of parametric confinement windows.  Let
\begin{equation*}
 \omega_{\rm loc}^2\equiv\frac{j(j+1)}{r_{\rm wall}^2}+\mu^2
\end{equation*}
denote the characteristic positive local mass/centrifugal scale near the
wall.  Modes with
\begin{equation}
 \sigma_{\rm in}\,\omega_{\rm loc}\;\lesssim\;\varpi\;\lesssim\;\omega_{\rm loc}
\label{eq:confinement-window}
\end{equation}
have classically allowed regions in the core and throat but are excluded from
the asymptotic exterior.  In physical terms, an excitation with an ordinary
local frequency as measured by an observer inside the de~Sitter core has,
after the redshift \(\sigma_{\rm in}\), a boundary frequency far below the
local mass and centrifugal scales of the AdS region and therefore cannot
propagate to the boundary.\footnote{
The last term in Eq.~\eqref{eq:schrodinger-form} should not in
general be omitted.  Defining
\begin{equation}
  {\cal C}(r)
  \equiv
  \frac{1}{r\sigma}\frac{d(\sigma N)}{dr},
\end{equation}
the Einstein equations give
\begin{equation}
  {\cal C}(r)
  =
  \frac{1-N}{r^2}
  +\frac{3}{L^2}
  +4\pi G(p_r-\varepsilon).
\end{equation}
Thus \({\cal C}\) has no inverse-\(N\) enhancement as long as the
local stress tensor remains controlled, and the full potential is
still suppressed by \({\cal Z}^2\) in a low-redshift region.
Nevertheless, \({\cal C}\) may be comparable to the mass and
centrifugal terms and must be included in determining the precise
turning points.  It vanishes at the minimum of the scalar-free
RN--AdS throat, equals \(-2H^2\) in the de Sitter core, and approaches
\(2/L^2\) near the AdS boundary.
}

Modes with still lower frequency,
\begin{equation}
 \mathcal{Z}_{\rm min}\,\omega_{\rm loc}\;\lesssim\;\varpi\;\lesssim\;
 \sigma_{\rm in}\,\omega_{\rm loc},
\label{eq:throat-window}
\end{equation}
are excluded from the central part of the core, where
\(\mathcal{Z}^2\simeq\sigma_{\rm in}^2\), and are concentrated in the
near-wall/throat region where \(\mathcal{Z}\) is smaller.  Since
\(\mathcal{Z}^2\) also decreases toward the outer part of the de~Sitter core,
these statements describe parametric localization windows rather than sharp
boundaries applying at every radius.  The deeper the throat, the lower the
frequencies it can support and the wider the second window becomes.  The
effective potential ultimately grows without bound toward the AdS boundary,
so the modes in both windows possess an outer turning point and are localized
global normal modes.  The location of that turning point, and hence the degree
of localization near the core and throat, depends on the detailed potential.

Raising the angular momentum does not remove the effect.  The centrifugal
contribution outside the wall and that inside the core are both proportional
to \(j(j+1)\), while \(r_{\rm core}\sim r_{\rm wall}\); their threshold ratio is
therefore controlled parametrically by \(\sigma_{\rm in}\), independently of
\(j\).  The same redshift argument applies directly to massless modes with large $j$, for which the centrifugal term supplies the local scale.  The
massless \(j=0,1\) modes are controlled by the curvature term in
Eq.~\eqref{eq:schrodinger-form} and must be examined separately.

For any fixed classical solution, the number of such localized levels is
finite.  With the physical tortoise length of the confining region
\begin{equation*}
 \mathcal{L}\equiv\int_{\rm core+throat}\frac{dr}{\sigma N},
\end{equation*}
the boundary-frequency spacing is
\(\Delta\varpi\sim\pi/\mathcal{L}\).  
Its growth toward the extremal limit, where \(\mathcal{L}\) diverges as \(1/\sqrt{N_{\min}}\), is discussed in Sec.~\ref{sec:extremal-limit}.

We emphasize that the existence of these localized global normal modes in
the free approximation does not rely on the wall acting as an elastic
reflector.  Localization is produced by the outer potential step, i.e.\
by the growth of \(V_j\) toward the asymptotically AdS region.  When the local
frequencies at the bottom of the throat reach the Planck scale and the wall
becomes absorptive rather than reflecting (see below), the potential analysis
continues to determine where the modes are (approximately) supported.

See also Refs.~\cite{Chen:2025cee,Chen:2025ywj,Chen:2025jbf} for the analysis of propagation of fields in the similar bulk geometry and its implication to the bulk-cone singularity and echoes appearing in the dual CFT correlator.
The model of the geometry, called a (AdS) gravastar, is constructed via Israel's junction condition and has a dS region inside, a AdS Schwarzschild region outside, and a thin shell at the junction surface.

\paragraph{As local physics, the core is nearly independent.}
The statement above concerns the global normal modes, defined with
respect to the boundary time.  Local physics inside the de~Sitter core looks
quite different, for a simple reason: the core is large in both space and time.
It is large in space---its proper size is the de~Sitter radius \(\sim1/H\),
many times the wall thickness---and it is large in time, since the near-extremal
throat separating it from the exterior has a long tortoise length, so a signal
produced in the core takes a parametrically long (boundary) time merely to
reach the throat.  Consequently, to a local, short-time probe---an observer
inside the core, over times short compared with the time to reach the
throat---the core simply looks like ordinary de~Sitter space at the scale
\(H\), with no reference to the near-extremality of the exterior.  The near-extremality
enters only when this local physics is translated to the boundary, where the
same core appears through strongly redshifted, closely spaced levels.

\paragraph{At the Planck scale, free modes lose meaning---but reconstruction does not.} As the wall is pushed to within a Planck length of the would-be
horizon, the blueshift $\omega_{\rm loc}=\varpi/(\sigma\sqrt N)$ at the throat
grows until the local gravitational coupling $G\,\omega_{\rm loc}^2$ becomes
$O(1)$. There the free normal-mode description breaks down: the wall acts not as
an elastic (Dirichlet) reflector but as a dissipative absorber, scattering
incident quanta into the dense bath of near-wall modes \cite{Terashima:2026fix, Terashima:2025tct}.
Consequently, the simple bulk
reconstruction of operators deep inside the core also becomes unclear in this
regime. 

\subsection{The extremal limit and the finite ${\mathcal N}$ black hole}
\label{sec:extremal-limit}

We now consider the limit toward extremality, in the spirit of the analysis of
\cite{Iizuka:2013kma} and
\cite{Terashima:2026fix, Terashima:2025tct, Chikazawa:2026iro}.\footnote{
Here, we assume that the string scale and the Planck scale are of the same order for simplicity.
}  

We work in a fixed holographic theory with a large but finite number
of degrees of freedom,
\begin{equation}
    \mathcal N^2 \equiv \frac{L^2}{G}\gg1,
\end{equation}
and with fixed bulk masses and couplings.  The parameters are chosen
in the controlled hierarchy
\begin{equation}
    \ell_{\rm Pl}
    \ll
    \Lambda_{\rm wall}^{-1}
    \sim m^{-1}
    \ll
    H^{-1}
    \lesssim L,
\end{equation}
or equivalently
\begin{equation}
    1\ll\frac{m}{H}\lesssim mL\ll \mathcal N.
\end{equation}
We then consider charged soliton states for which the RN--AdS
exterior is close to extremality and its metric function has a small
positive minimum,
\begin{equation}
    0<N_{\min}\ll1.
\end{equation}
The scalar field is already negligible in this exterior region, and
the boundary normalization gives \(\sigma=1\).  The redshift between
the boundary frame and the local frame near the minimum is therefore
\begin{equation}
    {\cal Z}_{\min}
    \simeq
    \sqrt{N_{\min}}
    \ll1.
\label{eq:minimum-redshift}
\end{equation}
The core remains of order \(H^{-1}\); the small quantity
\(N_{\min}\) characterizes the near-extremality of the exterior and
not a shrinking of the core.

The formal classical solution can be extrapolated to
\(N_{\min}=0\), where the exterior becomes extremal RN--AdS.  This
endpoint is not a physical limit at finite \(\mathcal N\).  Because
local energies near the minimum are enhanced by
\({\cal Z}_{\min}^{-1}\), the semiclassical free-field description
breaks down when the locally blueshifted modes reach the gravitational
cutoff.  The near-minimum region must then be replaced by an effective
brick wall or stretched horizon.

\paragraph{Finite-\(\mathcal N\) crossover and the quantum
completion of the throat.}
The exact classical extremal limit, with its infinitely deep throat, cannot
actually be taken at finite ${\mathcal N}$, as discussed in \cite{Terashima:2026fix}. For the generic black hole, the blueshifted local
energy near the stretched horizon reaches the Planck scale, and the classical/semiclassical
description is cut off there---this is the finite ${\mathcal N}$, brick-wall cutoff.\footnote{
Such a discrepancy between the finite ${\mathcal N}$ and large ${\mathcal N}$ expansion in AdS/CFT is also discussed in \cite{Sugishita:2022ldv, Sugishita:2023wjm, Sugishita:2024lee, Terashima:2025shl} from the perspective of bulk reconstruction.
}
Furthermore, for a near-extremal black hole, the Hawking temperature sets the
infrared scale at which nonextremality terminates the classical
near-horizon throat.  Finite-\(\mathcal N\) effects introduce a second infrared
scale of order \(S_{\rm BH}^{-1}\).  The crossover
\begin{equation}
  T_H \sim \frac{1}{r_h S_{\rm BH}} \sim
  \frac{1}{{\mathcal N}^{2}} \frac{L^2}{(r_h)^3},
  \label{eq:TH-scaling}
\end{equation}
therefore marks the point at which the finite-\({\mathcal N}\) cutoff becomes
relevant before the formal semiclassical throat reaches its
near-horizon endpoint.  
Here, $S_{\rm BH}$ is the entropy of the corresponding extremal black hole
and we used the results in \cite{Terashima:2026fix} slightly modified form because $L\sim r_h$ was assumed in \cite{Terashima:2026fix}.

Below
this scale  \eqref{eq:TH-scaling},  the smooth near-extremal geometry ceases to provide a
controlled description of the finite-\(\mathcal N\) state.
Thus, the formal classical thin-wall limit should not be identified with a
smooth solution at \(N_{\min}=0\).  
The appropriate finite-\(\mathcal N\)
description is therefore not a colder classical soliton, but a
regular de Sitter core capped by an interacting stretched-horizon
sector, with an RN--AdS-like exterior.
The stretched-horizon
fluctuations carry an entropy of order \(A_h/4G\) and provide the
brick-wall completion of the classical thin-wall configuration, as discussed in \cite{Terashima:2026fix}.  
The
formal continuation \(N_{\min}\to0\) only indicates the direction in
which the classical solution approaches this finite-\({\mathcal N}\) regime.

\paragraph{The black-hole low-energy degrees of freedom and the de Sitter core.}
In the parametrically thin-wall limit
described in Sec.~\ref{sec:fuzzball-buchdahl}, the soliton wall is expected to
approach both (i) the de~Sitter cosmological horizon, at \(r\simeq H^{-1}\),
and (ii) the would-be horizon of the nearly extremal black hole, at
\(r\simeq r_h\).  Thus the limiting gluing gives \(H^{-1} = r_h\), up to $1/{\mathcal N}$ corrections.  The
explicit solutions found so far exhibit the same hierarchy of macroscopic
scales, although they do not yet reach this limiting coincidence.  On the
limiting surface, the screen viewed from outside is the brick-wall screen of
the near-extremal black hole, while viewed from inside it is the stretched
horizon of the de~Sitter static patch.\footnote{%
It is tempting to speculate that this points to a connection with
the de Sitter static patch holography, in which the observables of a static
patch are encoded on a screen placed at (or near) its stretched horizon
\cite{Shaghoulian:2022fop, Susskind:2023rxm}. In the present construction that
screen is not a separate, postulated boundary but a concrete bulk
surface---the soliton wall---which simultaneously serves as the brick wall of
the dual (asymptotically AdS) black hole in the limiting geometry. The de
Sitter static patch is thereby embedded inside an ordinary asymptotically AdS
spacetime with a genuine holographic dual. Whether this identification can be
made sharp---in particular, whether the static-patch screen degrees of
freedom are precisely the brick-wall/low-energy sector described here, and
how the de Sitter (false-vacuum) data are encoded in the charge-$Q$ ground state---is left as an open question.}
In the limiting
common-surface identification, the de~Sitter stretched horizon and the
black-hole brick wall enclose the (almost) same area,
\begin{equation}
 A=4\pi r_h^2= 4\pi H^{-2}.
\end{equation}

The corresponding entropy counts have the same area-law scaling.  The
de~Sitter static patch has entropy \(S_{\rm dS}=A/(4G)\), and the nearly
extremal black hole seen from outside has
\(S_{\rm BH}=A/(4G)\) for the same area.  The brick-wall count of modes
localized just inside the wall, with a proper-distance cutoff of order
\(\ell_{\rm Pl}\), gives
\(S_{\rm BW}\sim A/\ell_{\rm Pl}^2\sim A/G\).\footnote{
The precise numerical
coefficient in the brick-wall expression depends on the regulator and field
content and is understood together with the usual renormalization of
Newton's constant.}  
Thus all three entropies scale as
\begin{equation}
 S\sim\frac{A}{G}.
\label{eq:entropy-identification}
\end{equation}
The agreement follows because a single limiting surface, with a single area
and a single proper cutoff, is being described in three languages.  
Therefore, in our setting,
the black hole and the
de~Sitter entropies have the same origin in the brick-wall degrees of freedom around the (common) stretched horizon.
Since the full soliton is horizonless, \(S_{\rm dS}\) and
\(S_{\rm BH}\) here do not represent two independent horizon
entropies.  They are two descriptions, from the interior and exterior
sides, of the entropy carried by the same quantum fluctuations
localized near the common would-be horizon.  The three counts
\(S_{\rm dS}\), \(S_{\rm BH}\), and \(S_{\rm BW}\) should therefore
not be added; they refer to the same near-wall degrees of freedom.
Their equality is consequently stronger than a coincidence of
area-law scaling: within the semiclassical brick-wall description,
the same fluctuation sector is being counted in three different
geometric languages.\footnote{
Here entropy means the coarse-grained number of quantum states
associated with the semiclassical background \cite{Terashima:2025shl}, rather than the
von Neumann entropy of a single exact pure CFT state.
}

\paragraph{CFT operators probing the de Sitter core.}
The bulk hierarchy described above suggests a direct way to access the
de Sitter core from the CFT.  Consider a bulk operator
\(\Phi_{\rm core}\) supported in a compact region well inside the core
and away from the wall.  The very-low-energy modes responsible for the
stretched-horizon entropy are localized near the wall and have
parametrically small support in this interior region.  Consequently,
\(\Phi_{\rm core}\) is insensitive to these entropy-dominating modes
at leading order, while it couples to the higher, but still strongly
redshifted, modes that penetrate the de Sitter core.

If the bulk background constructed here is realized in AdS/CFT,
\(\Phi_{\rm core}\) must admit a boundary representation, which we
denote by \(\mathcal O_{\rm core}\).  Since the proper static-patch
time in the core is related to the boundary time by
\begin{equation}
    \tau \simeq \sigma_{\rm in} t,
\end{equation}
a core excitation with local energy of order \(H\) appears in the
boundary Hamiltonian at the low-energy scale
\begin{equation}
    E-E_Q \sim \sigma_{\rm in} H,
\end{equation}
where $E_Q$ is the energy of the lowest energy state with charge $Q$.
Thus, \(\mathcal O_{\rm core}\) should be accessible in a
low-energy, long-time regime of the CFT and should probe the de Sitter
core without resolving the entropy-dominating stretched-horizon
degrees of freedom.  No separate factorization or detailed assumption
about the CFT spectrum is required for this statement; the separation
follows from the localization properties of the corresponding bulk
modes.

What is not presently known is the explicit construction of
\(\mathcal O_{\rm core}\).  Semiclassically, the boundary
representation is expected to be highly nonlocal in time because a
signal from the core must traverse the long near-extremal throat.
Moreover, near the finite-\({\mathcal N}\) cutoff the wall becomes strongly
coupled and absorptive, so the simple free-field reconstruction loses
control.  We therefore expect a low-energy CFT representation of
operators in the de Sitter core to exist if this semiclassical bulk
setting has a holographic realization, but its concrete construction
and intrinsic CFT characterization remain open.

\paragraph{Which de Sitter state?}
The full CFT state is stationary, but stationarity by itself does not
select a de~Sitter vacuum.  The relevant additional input is the
semiclassical brick-wall interpretation adopted above.  The quantum
fluctuations localized near the common stretched-horizon surface form
a thermal atmosphere.  In the common-horizon limit, this is the same
fluctuation sector viewed, from the exterior, as the black-hole
brick-wall sector and, from the interior, as the de~Sitter
stretched-horizon sector.

With
\begin{equation}
  \tau\simeq\sigma_{\rm in}t
\end{equation}
the proper static time near the center of the core, the
Gibbons--Hawking temperature is
\begin{equation}
  T_{\rm dS}^{(\tau)}=\frac{H}{2\pi},
  \qquad
  T_{\rm dS}^{(t)}
  =\frac{\sigma_{\rm in}H}{2\pi}.
\end{equation}
In the common-horizon limit, regularity requires the near-wall sector
to exhibit the corresponding KMS behavior at this temperature, at
leading semiclassical order.  Correlators of operators supported a
fixed proper distance inside the core then approach those of the
Bunch--Davies vacuum restricted to a static patch, up to wall-induced
corrections and, at sufficiently long times, reflected signals from
the wall.  At any finite wall position there is no actual horizon, so
the would-be Boulware singularity is cut off by the wall; in the
common-horizon limit, the thermal KMS state at the
Gibbons--Hawking temperature gives the regular limiting behavior.

A global Bunch--Davies state is not realized, because the global
de~Sitter geometry and the second static patch are absent.  They are
replaced by the wall and the near-extremal throat, which play the role
of the complementary sector whose entanglement produces the thermal
static-patch state.  This statement does not require an exact
factorization of the gravitational Hilbert space.  The classification
by exact de~Sitter-invariant \(\alpha\)-vacua does not directly apply
at finite wall position, since the wall breaks the de~Sitter
isometries.  In the limiting de~Sitter region, local Euclidean
regularity of the core correlators selects the Bunch--Davies behavior
rather than a non-Euclidean \(\alpha\)-vacuum.

\section*{Acknowledgements}

ST would like to thank T. Kawamoto and K. Suzuki for useful discussions.
KM would like to thank H. Ishihara for valuable discussions.
This work was supported by JSPS KAKENHI Grant Number 24K07048(ST) and 23KK0048(YK).

\paragraph*{Declaration of Generative AI Assistance}
The authors retain full responsibility for the concepts, methodology, and narrative of this manuscript. Generative AI tools, specifically Claude and OpenAI models, assisted with aspects of the analysis, proofreading, text formatting, and sanity-checking specific quantitative calculations.

\appendix

\section{AdS Ogawa--Ishihara model with a weakly gauged relative U(1)}
\label{sec:OI}

This section gives the AdS version of the two-field model of Ogawa--Ishihara,
with one additional weak gauge field for the U(1) symmetry that is global in the original model.
The original model contains two complex scalars, which we denote by $\psi$ and $\varphi$.
The scalar $\varphi$ is the Higgs field with symmetry-breaking scale $\eta$, while $\psi$ is the matter/Q-ball field.
The potential is
\begin{equation}
U(|\varphi|,|\psi|)
=\frac{\lambda}{4}\left(|\varphi|^2-\eta^2\right)^2
+\mu_{\psi\varphi}|\psi|^2|\varphi|^2.
\label{eq:OI-potential}
\end{equation}
We now include a negative cosmological constant and gauge both U(1)'s.
The first gauge field $A_\mu$ is the ordinary gauge field already present in the Ogawa--Ishihara model.
The second gauge field $B_\mu$ weakly gauges the relative U(1) symmetry.
We take its coupling $q$ to be small.

\paragraph{Action and conventions.}

We use the same metric convention as in the preceding sections and write
\begin{equation}
\Lambda=-\frac{3}{L^2},
\qquad
F_{\mu\nu}=\partial_\mu A_\nu-\partial_\nu A_\mu,
\qquad
H_{\mu\nu}=\partial_\mu B_\nu-\partial_\nu B_\mu.
\end{equation}
The action is
\begin{align}
S_{\rm OI,AdS}
=\int d^4x\sqrt{-g}\Bigg[
&\frac{1}{16\pi G}\left(R-2\Lambda\right)
-\frac14 F_{\mu\nu}F^{\mu\nu}
-\frac14 H_{\mu\nu}H^{\mu\nu} 
\nonumber\\
&-(D_\mu\psi)^*(D^\mu\psi)
-(D_\mu\varphi)^*(D^\mu\varphi)
-U(|\varphi|,|\psi|)
\Bigg].
\label{eq:OI-AdS-action}
\end{align}
We choose charges so that $A_\mu$ gauges the diagonal U(1), while $B_\mu$ gauges the relative U(1):
\begin{equation}
D_\mu\psi=(\nabla_\mu+i e A_\mu-iq B_\mu)\psi,
\qquad
D_\mu\varphi=(\nabla_\mu+i e A_\mu+iq B_\mu)\varphi.
\label{eq:OI-covder}
\end{equation}
The limit $q\to0$ returns to the case in which the relative U(1) is global.
The asymptotically flat version with $q=0$ and $L\to\infty$ is the original Ogawa--Ishihara system, up to the sign convention for the gauge potential.

\paragraph{Spherically symmetric ansatz.}

We use
\begin{equation}
 ds^2=-\sigma(r)^2N(r)dt^2+\frac{dr^2}{N(r)}+r^2d\Omega_2^2,
 \qquad
 N(r)=1-\frac{2m(r)}{r}+\frac{r^2}{L^2}.
\label{eq:OI-AdS-metric}
\end{equation}
For the matter fields, we take
\begin{equation}
\psi(t,r)=u(r)e^{-i\omega_\psi t},
\qquad
\varphi(t,r)=f(r)e^{-i\omega_\varphi t},
\qquad
A=A_0(r)dt,
\qquad
B=B_0(r)dt.
\label{eq:OI-ansatz}
\end{equation}
The gauge-invariant local frequencies are
\begin{equation}
\Omega_\psi(r)=\omega_\psi-eA_0(r)+qB_0(r),
\qquad
\Omega_\varphi(r)=\omega_\varphi-eA_0(r)-qB_0(r).
\label{eq:OI-Omegas}
\end{equation}
If both U(1)'s are gauged, one may use gauge transformations to remove the two time-dependent phases.
Keeping $\Omega_\psi$ and $\Omega_\varphi$ explicit is often more convenient for comparison with the one-field equations above.

\paragraph{Radial equations of motion.}

The energy density entering $m'(r)$ is
\begin{align}
\varepsilon
=&\frac{A_0'^2+B_0'^2}{2\sigma^2}
+\frac{\Omega_\psi^2u^2+\Omega_\varphi^2f^2}{\sigma^2N}
+N\left(u'^2+f'^2\right)
+U(f,u),
\label{eq:OI-energy-density}
\end{align}
where
\begin{equation}
U(f,u)=\frac{\lambda}{4}\left(f^2-\eta^2\right)^2
+\mu_{\psi\varphi}u^2f^2.
\end{equation}
The Einstein equations reduce to
\begin{align}
m'(r)&=4\pi G r^2\varepsilon,
\label{eq:OI-mprime}\\
\frac{\sigma'(r)}{\sigma(r)}
&=\frac{8\pi G r}{N(r)}
\left[
\frac{\Omega_\psi^2u^2+\Omega_\varphi^2f^2}{\sigma^2N}
+N\left(u'^2+f'^2\right)
\right].
\label{eq:OI-sigmaprime}
\end{align}
The scalar equations are
\begin{align}
\frac{1}{\sigma r^2}\frac{d}{dr}\left(\sigma r^2N u'\right)
&=\mu_{\psi\varphi} f^2u
-\frac{\Omega_\psi^2}{\sigma^2N}u,
\label{eq:OI-u-eq}\\
\frac{1}{\sigma r^2}\frac{d}{dr}\left(\sigma r^2N f'\right)
&=\frac{\lambda}{2}f\left(f^2-\eta^2\right)
+\mu_{\psi\varphi}u^2f
-\frac{\Omega_\varphi^2}{\sigma^2N}f.
\label{eq:OI-f-eq}
\end{align}
The two Gauss-law equations are
\begin{align}
\frac{1}{\sigma r^2}\frac{d}{dr}\left(\frac{r^2 A_0'}{\sigma}\right)
&=-\frac{2e}{\sigma^2N}\left(\Omega_\psi u^2+\Omega_\varphi f^2\right),
\label{eq:OI-A-eq}\\
\frac{1}{\sigma r^2}\frac{d}{dr}\left(\frac{r^2 B_0'}{\sigma}\right)
&=\frac{2q}{\sigma^2N}\left(\Omega_\psi u^2-\Omega_\varphi f^2\right).
\label{eq:OI-B-eq}
\end{align}
The signs follow from the charge assignments in \eqref{eq:OI-covder}.
Equivalently, the equations can be written directly in terms of the two gauge-invariant combinations \eqref{eq:OI-Omegas}.

\paragraph{Dimensionless form.}

For this model the natural mass scale is the Higgs scale $\eta$.
We define
\begin{equation}
x\equiv \eta r,
\qquad
\mathcal{M}(x)\equiv \eta m(r),
\qquad
\ell\equiv \eta L,
\qquad
\alpha\equiv 4\pi G\eta^2,
\end{equation}
\begin{equation}
s(x)\equiv \frac{u(r)}{\eta},
\qquad
h(x)\equiv \frac{f(r)}{\eta},
\qquad
 a(x)\equiv \frac{eA_0(r)}{\eta},
\qquad
 b(x)\equiv \frac{qB_0(r)}{\eta},
\end{equation}
\begin{equation}
\varpi_\psi\equiv \frac{\omega_\psi}{\eta},
\qquad
\varpi_\varphi\equiv \frac{\omega_\varphi}{\eta},
\qquad
W_\psi\equiv \varpi_\psi-a+b,
\qquad
W_\varphi\equiv \varpi_\varphi-a-b.
\end{equation}
Then
\begin{equation}
N(x)=1-\frac{2\mathcal{M}(x)}{x}+\frac{x^2}{\ell^2},
\qquad
\widetilde U(h,s)=\frac{\lambda}{4}(h^2-1)^2+\mu_{\psi\varphi}s^2h^2,
\end{equation}
where a prime now denotes $d/dx$.
The dimensionless Einstein equations are
\begin{align}
\mathcal{M}'
=\alpha x^2\Bigg[
&\frac{a'^2}{2e^2\sigma^2}
+\frac{b'^2}{2q^2\sigma^2}
+\frac{W_\psi^2s^2+W_\varphi^2h^2}{\sigma^2N}
+N(s'^2+h'^2)
+\widetilde U(h,s)
\Bigg],
\label{eq:OI-dimless-M}\\
\frac{\sigma'}{\sigma}
=\frac{2\alpha x}{N}
\left[
\frac{W_\psi^2s^2+W_\varphi^2h^2}{\sigma^2N}
+N(s'^2+h'^2)
\right].
\label{eq:OI-dimless-sigma}
\end{align}
The scalar equations are
\begin{align}
\frac{1}{\sigma x^2}\frac{d}{dx}\left(\sigma x^2Ns'\right)
&=\mu_{\psi\varphi}h^2s
-\frac{W_\psi^2}{\sigma^2N}s,
\label{eq:OI-dimless-s}\\
\frac{1}{\sigma x^2}\frac{d}{dx}\left(\sigma x^2Nh'\right)
&=\frac{\lambda}{2}h(h^2-1)
+\mu_{\psi\varphi}s^2h
-\frac{W_\varphi^2}{\sigma^2N}h.
\label{eq:OI-dimless-h}
\end{align}
Finally, the two corrected Gauss-law equations are
\begin{align}
\frac{1}{\sigma x^2}\frac{d}{dx}\left(\frac{x^2a'}{\sigma}\right)
&=-\frac{2e^2}{\sigma^2N}\left(W_\psi s^2+W_\varphi h^2\right),
\label{eq:OI-dimless-a}\\
\frac{1}{\sigma x^2}\frac{d}{dx}\left(\frac{x^2b'}{\sigma}\right)
&=\frac{2q^2}{\sigma^2N}\left(W_\psi s^2-W_\varphi h^2\right).
\label{eq:OI-dimless-b}
\end{align}
These are the direct two-field analogues of the corrected one-field equation \eqref{eq:dS-w}.
The weak-gauging limit of the relative U(1) is $q\ll e$ or simply $q\ll1$, depending on the normalization of the first gauge coupling.

\paragraph{Boundary conditions and dS-like core.}

Regularity at the origin requires
\begin{equation}
\mathcal{M}(0)=0,
\qquad
s'(0)=h'(0)=a'(0)=b'(0)=\sigma'(0)=0.
\end{equation}
For an asymptotically AdS gravastar-like soliton, one imposes
\begin{equation}
s(\infty)=0,
\qquad
h(\infty)=1,
\qquad
\sigma(\infty)=1,
\end{equation}
with vanishing field strengths at infinity.
The finite-energy Higgs condition is
\begin{equation}
W_\varphi(\infty)=0,
\end{equation}
up to a residual gauge choice.
If the exterior is to be neutral under both gauge fields, one further imposes
\begin{equation}
a'(\infty)=b'(\infty)=0.
\end{equation}
The Ogawa--Ishihara gravastar core corresponds to
\begin{equation}
h\simeq0,
\qquad
s\simeq s_0,
\qquad
W_\psi\simeq0,
\end{equation}
so that the dominant energy density is the false-vacuum energy
\begin{equation}
\varepsilon_0\simeq \frac{\lambda}{4}\eta^4.
\end{equation}
In asymptotically AdS, a necessary condition for the core to be locally dS-like is
\begin{equation}
-\frac{3}{L^2}+8\pi G\varepsilon_0>0,
\qquad\text{i.e.}\qquad
\frac{\alpha\lambda}{2}>\frac{3}{\ell^2}.
\label{eq:OI-dS-condition-dimless}
\end{equation}
This condition is only necessary: the full solution also requires the wall fields to interpolate to
$s\to0$, $h\to1$ without creating a horizon or a singularity.

\section{Asymptotically flat solutions}
\label{sec:dS-RN-largeL}

In this appendix, we focus on the asymptotically flat case, $\ell=\infty$, and demonstrate that regular solutions with a de Sitter core also exist in this limit. In this case, our setup is identical to that considered in \cite{Brihaye:2025dlq}. 
The numerical solutions presented below are obtained using parameter regions that differ from those in \cite{Brihaye:2025dlq}.

\begin{figure}
  \centering
\subfigure[]
 {\includegraphics[scale=0.6]{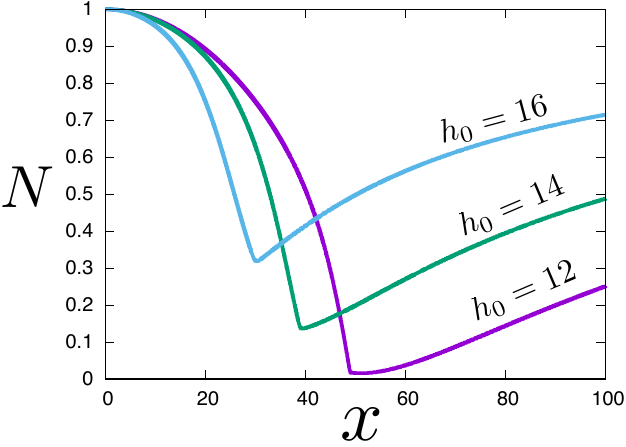}\label{N_f}
  }
  \subfigure[]
 {\includegraphics[scale=0.6]{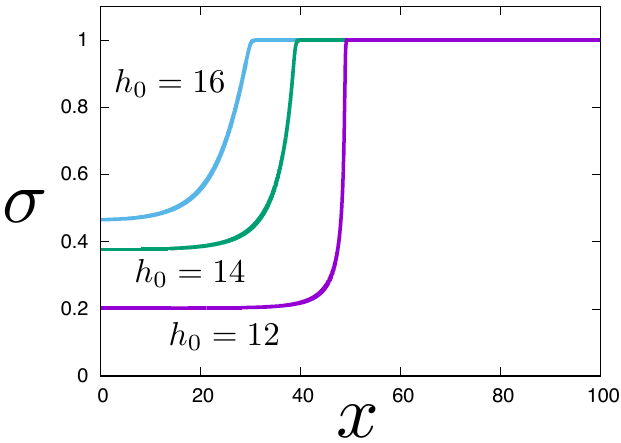}\label{sigma_f}
  }
   \subfigure[]
 {\includegraphics[scale=0.6]{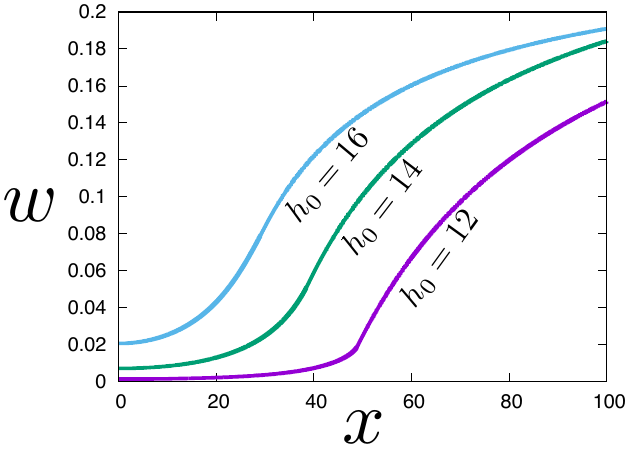}\label{w_f}
  }
  \subfigure[]
 {\includegraphics[scale=0.6]{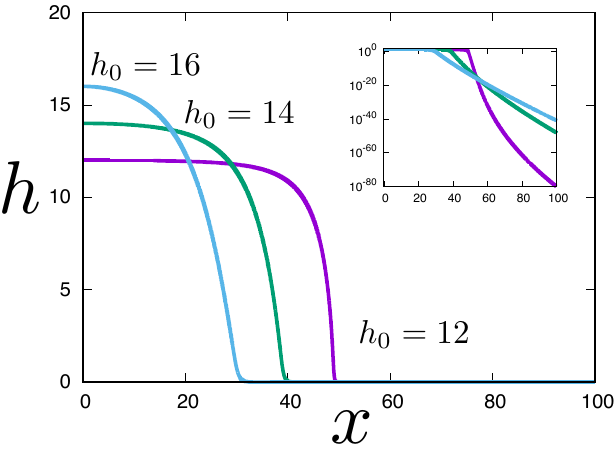}\label{h_f}
  }
 \caption{
 Numerical solutions for $N$, $\sigma$, $w$, and $h$ in the asymptotically flat case. The parameters are fixed to the values given in Eq.~(\ref{eq:paramset-L}). 
  The scalar field at the origin is varied as $h_0=12$, $14$, and $16$. Although the numerical integration is performed up to $x=200$, only the region $x\le100$ is shown for clarity. The inset in panel (d) shows the scalar field on a logarithmic scale.
}
 \label{sols_f}
\end{figure}

\begin{figure}
  \centering
  \subfigure[Energy density]
 {\includegraphics[scale=0.45]{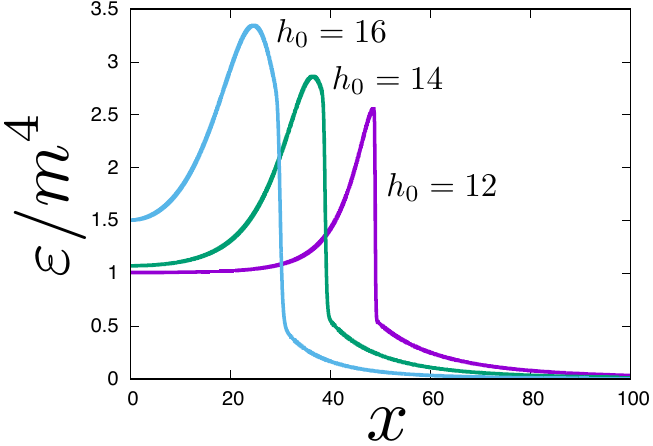}\label{energy_f}
  }
  \subfigure[Radial pressure]
 {\includegraphics[scale=0.45]{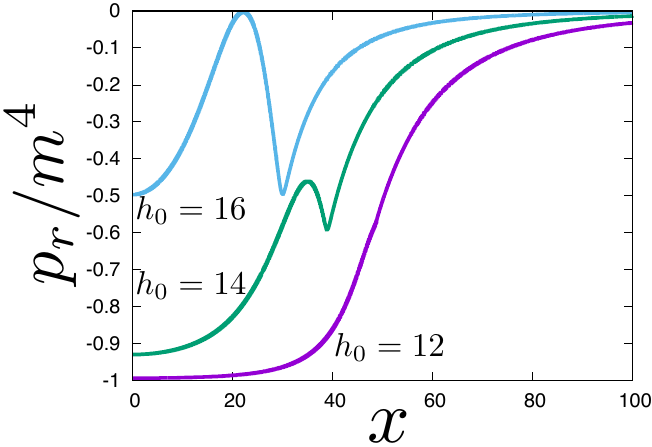}\label{pr_f}
  }
  \subfigure[Tangential pressure]
 {\includegraphics[scale=0.45]{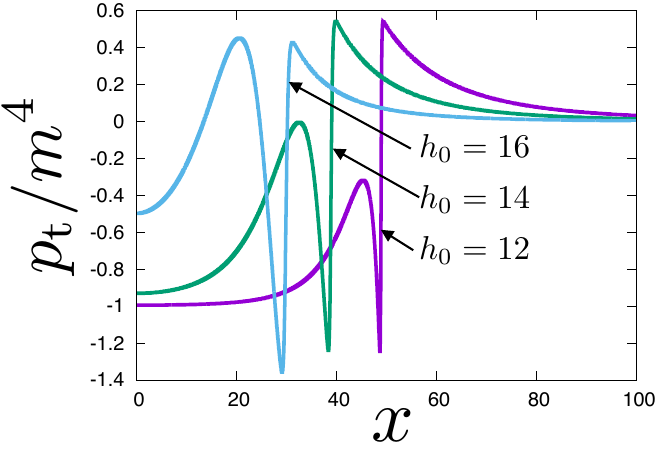}\label{pt_f}
  }
 \caption{Energy density, radial pressure, and tangential pressure in the asymptotically flat case. The parameters are fixed to the values given in Eq.~(\ref{eq:paramset-L}).  The scalar field at the origin is varied as $h_0=12$, $14$, and $16$.
}
 \label{energy_pressure_f}
\end{figure}

The parameters are chosen as 
\begin{equation}
  \tilde\eta=1,\quad
  \alpha=4\times10^{-4},\quad
  \ell=\infty,\quad
  e=5\times10^{-3}. 
  \label{eq:paramset-L}
\end{equation}
The regular expansions near the origin, Eqs.~(\ref{expand0o}) and (\ref{r0expand}), and at spatial infinity, Eqs.~(\ref{eq:bc-h-Q}) and (\ref{expandinf2}), 
are obtained essentially by taking the naive limit $\ell\to\infty$ of the asymptotically AdS case. The only exception is the asymptotic behavior of the scalar field, 
which is replaced by the decaying massive tail,
\begin{equation}
h(x)
=
h_\infty
\left(\frac{x}{x_{\rm max}}\right)^{\beta}
\exp\!\left[
-\kappa\,(x-x_{\rm max})
\right],
\,\,
\kappa=\sqrt{1-w_\infty^2},
\,\,
\beta=-1+\frac{w_\infty c_w+M\,(2w_\infty^2-1)}{\kappa},
\label{eq:bc-h-flat}
\end{equation}
with $c_w=-eQ/\sqrt{\alpha}$.
The power $\beta$ comes from the $1/x$ terms of $N$ and $w$ in Eq.~(\ref{expandinf2}), i.e., from the Newtonian potential of $M$ and the Coulomb potential of $Q$ acting on the massive scalar; for $M=c_w=0$ it reduces to $\beta=-1$.
Figure~\ref{sols_f} shows the numerical solutions for $h_0=12$, $14$, and $16$. Figure~\ref{energy_pressure_f} shows the energy density, radial pressure, and tangential pressure. For readers who wish to reproduce our numerical calculation, we provide the converged values of these parameters for the case $h_0=12$ as
\begin{equation}
\begin{split}
&w_0=6.82968\times 10^{-3},\quad M=50.3378,\quad Q=50.7366 ,\\
&w_\infty=0.278284,\quad  \log_{10} h_\infty=-145.028,
\end{split}
\end{equation}
The numerical domain is specified by $x_{\rm min}=10^{-3}, x_{\rm mid}=48.8351, x_{\rm max}=200$.

As in the asymptotically AdS case, we obtain globally regular, horizonless solutions. 
For $x\lesssim x_{\rm wall}$, where $x_{\rm wall}\simeq49,\;39,\;29$ for $h_0=12,\;14,\;16$, respectively, the scalar field remains in the false vacuum, whose nonzero vacuum energy gives rise to a de Sitter--like core. Accordingly, the metric function $N(x)$ is well approximated by
$N(x)\simeq1-x^2/\ell_{\rm dS}^2$. 
Near $x\simeq x_{\rm wall}$, the scalar field rapidly decays to a small value. Correspondingly, the spacetime undergoes a transition from the de Sitter core to an almost asymptotically flat Reissner--Nordström geometry. For $x\gtrsim x_{\rm wall}$, the solution is therefore well approximated by the asymptotically flat Reissner--Nordström spacetime.

\begin{figure}
  \centering
  \subfigure[Mass–charge relation]
 {\includegraphics[scale=0.55]{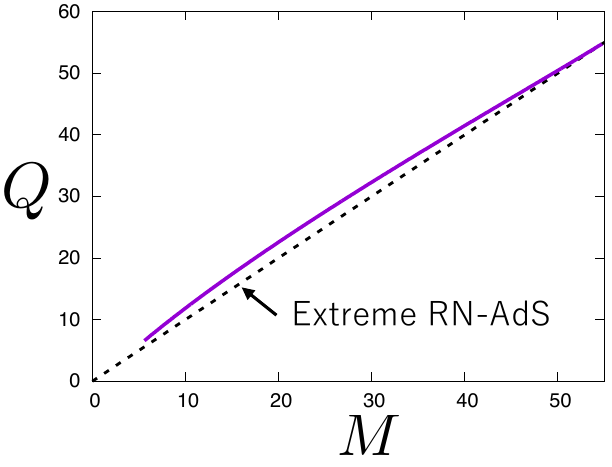}\label{MQ_f}
  }
  \subfigure[de Sitter radius and wall radius]
 {\includegraphics[scale=0.55]{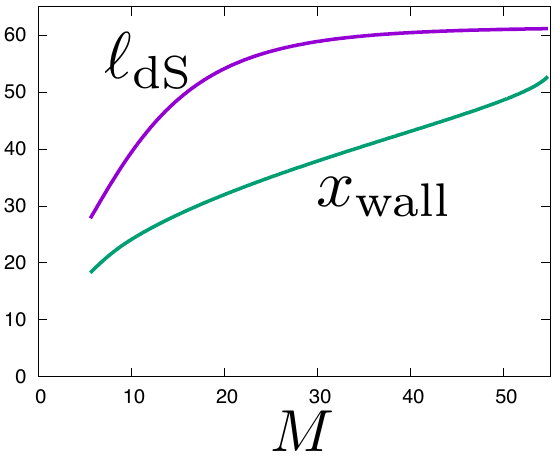}\label{Mlds_f}
  }
 \caption{The family of solutions obtained by varying $h_0$ in the asymptotically flat case. Panel (a) shows the solutions in the $M$--$Q$ plane. Panel (b) shows the de Sitter radius $\ell_{\rm dS}$ and the wall position $x_{\rm wall}$ as functions of the mass $M$. The parameters are fixed to the values given in Eq.~(\ref{eq:paramset-L}). 
}
 \label{MQlds_f}
\end{figure}

Figure~\ref{MQlds_f} shows the family of asymptotically flat solutions obtained with the model parameters fixed as in Eq.~(\ref{eq:paramset-L}). Panel (a) presents the $M$--$Q$ relation. Panel (b) shows the de Sitter radius $\ell_{\rm dS}$ and the wall position $x_{\rm wall}$ as functions of the mass $M$. The dashed curve in panel corresponds to the extremal Reissner--Nordström solution $M=Q$.
All solutions found in the present search lie in the superextremal region of the Reissner--Nordström as in the case of the asymptotically AdS.

\hspace{1cm}

\bibliographystyle{utphys}
\bibliography{main202506.bib}

\end{document}